\documentclass[prb,superscriptaddress,a4paper,floatfix,nofootinbib]{revtex4}

\usepackage{graphicx}
\usepackage{amssymb}
\usepackage{amsmath}
\usepackage{subfigure}     
\usepackage[usenames,dvipsnames]{color}
\usepackage{hhline}
\usepackage[T1]{fontenc}
\usepackage{moreverb}
\usepackage{braket}
\makeindex
\begin{document}

\title{Quantum magnetism with ultracold molecules}

\newcommand{\Notes}[1]{{}}
\newcommand{\todo}[1]{{\color{red}#1}}
\newcommand{\warn}[1]{{\color{red}\textbf{* #1  *}}}
\newcommand{\expec}[1]{\left < #1\right >}
\newcommand{\warncite}[1]{{\color{blue}\textbf{ *** CITE:  #1  *** }}}
\newcommand{\MF}{CH$_3$F}
\newcommand{\citen}{\onlinecite}
\newcommand{\tbl}{\caption}

\author{M. L. Wall }
\affiliation{JILA, NIST, and Department of Physics, University of Colorado, Boulder, CO 80309}

\author{K.~R.~A.~Hazzard}
\affiliation{JILA, NIST, and Department of Physics, University of Colorado, Boulder, CO 80309}

\author{Ana Maria  Rey}
\affiliation{JILA, NIST, and Department of Physics, University of Colorado, Boulder, CO 80309}



\begin{abstract}
This article gives an introduction to the realization of effective quantum magnetism with ultracold molecules in an optical lattice, reviews experimental and theoretical progress, and highlights future opportunities opened up by ongoing experiments. Ultracold molecules offer capabilities that are otherwise difficult or impossible to achieve in other effective spin systems, such as long-ranged spin-spin interactions with controllable degrees of spatial and spin anisotropy and favorable energy scales.  Realizing quantum magnetism with ultracold molecules provides access to rich many-body behaviors, including many exotic phases of matter and interesting excitations and dynamics.  Far-from-equilibrium dynamics plays a key role in our exposition, just as it did in recent ultracold molecule experiments realizing effective quantum magnetism.  In particular, we show that dynamical probes allow the observation of correlated many-body spin physics, even in polar molecule gases that are not quantum degenerate.  After describing how quantum magnetism arises in ultracold molecules and discussing recent observations of quantum magnetism with polar molecules, we survey prospects for the future, ranging from immediate goals to long-term visions. 
\end{abstract}

\date{\today}
\maketitle

\tableofcontents


\section{Introduction}
\label{sec:Intro}

The realization of a Bose-Einstein condensate (BEC) in an ultracold, dilute gas of alkali atoms~\cite{anderson1995,davis1995,bradleyCC1995} was a landmark achievement in several respects.  For one, the production of an atomic BEC required significant technical advances in cooling and trapping atoms with electromagnetic radiation, as well as evaporative cooling.  In addition, atomic BECs provided the most direct evidence for a many-body phenomenon predicted more than 80 years prior.  The fact that nearly all aspects of the atomic system are amenable to experimental control not only enabled the realization of this new state of matter, but also facilitated the study of its properties out of equilibrium, such as the dynamics of vortices~\cite{PhysRevLett.83.2498} and solitons~\cite{PhysRevLett.86.2926}.  As we shall see, ultracold molecules parallel this, providing entirely new phases of matter and non-equilibrium behaviors that are otherwise unrealized in ultracold matter.

Since the realization of a BEC the field of many-body physics with ultracold atoms has grown steadily~\cite{Bloch}, including degenerate fermionic gases~\cite{DeMarco} as well as many non-alkali species~\cite{AEAs,PhysRevLett.108.215301,PhysRevLett.107.190401,PhysRevLett.108.210401}.  A burgeoning subfield of research involves optical lattices, standing wave arrangements of laser light that form a periodic potential for atoms or molecules~\cite{bloch2005ultracold}.  Such a periodic potential mimics the crystal potential felt by electrons in a solid, enabling the atoms to "simulate" the behavior of interacting electrons in a crystal lattice.  The power of the atom-electron analogy comes from the fact that characteristics of the atomic system, such as lattice geometry, degree of disorder, and strength of interactions, are all highly tunable.  This enables the atoms to behave as a \emph{quantum simulator}, a quantum system that behaves analogously to another system (which may be much harder to microscopically control or measure)~\cite{feynmanRP1982}, and has led to many spectacular observations, such as the transition from a weakly-interacting gas to a Mott insulator for both bosonic~\cite{Greiner_Mandel_02} and fermionic~\cite{jordens2008mott,schneider:metallic_2008} atoms.  

In spite  of the successes of ultracold atom experiments, some phenomena remain difficult to manifest and probe in cold atoms.  One such phenomenon is quantum magnetic interactions between effective spins.  Quantum magnetism, which studies the many-body physics of coupled spins, is of key importance in condensed matter physics (see Sec.~\ref{sec:PartII}).  In atomic realizations of quantum magnetism, usually the "spin" is formed from some discrete set of internal states, for example hyperfine sublevels.  The reason why realizing effective magnetic interactions between such spins is difficult, as Sec.~\ref{sec:superexchange} describes in more detail, is that the dominant interactions between neutral atoms are short ranged; this requires any effective spin-spin interactions between internal states to be mediated by motion.  Hence, magnetic correlations  become visible only when the motional temperature is less than the effective spin-spin coupling energy.  To date, such temperatures are extraordinarily difficult to reach.

\begin{figure}[htp]
\centerline{\includegraphics[width=0.666\columnwidth]{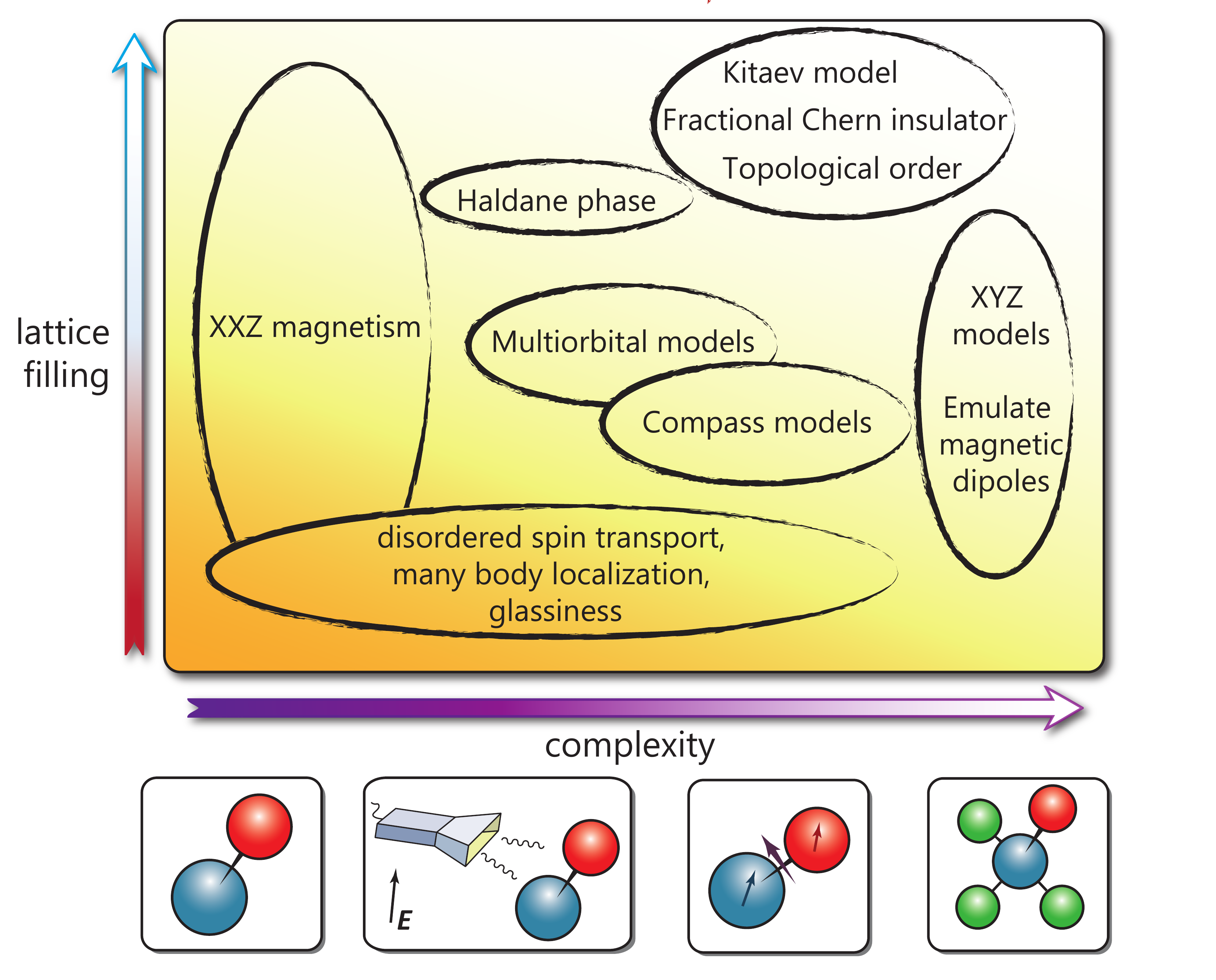}}
\caption{\label{fig:complexity}  Regimes of quantum magnetism in ultracold polar molecules in optical lattices, encompassing ongoing experiments (lower left) and directions being pursued (upper right).  The vertical "lattice filling" axis represents the fraction of occupied lattice sites, that is the molecule density relative to the lattice spacing. Higher fillings generally correspond to lower motional entropies and temperatures. Current experiments reach filling fractions in the range of  $10\%$-$20\%$.  Interesting physics exists in this regime, but exciting prospects also occur as one increases the filling fraction towards unity.  The horizontal  "complexity" axis represents complexity in two senses: the molecules' degrees of freedom and the experimental requirements to harness them.  Diverse, rich, and novel phenomena at the forefront of modern quantum many-body physics occur in all of the indicated regimes. 
}
\end{figure}

Ultracold molecules are a newly realized platform for quantum magnetism in which the magnetism arises in a qualitatively distinct way from atoms, and  this underlies the molecules' many favorable qualities.  
As opposed to atoms, polar molecules have strong, long-range electric dipole-dipole interactions\footnote{Homonuclear molecules are not polar, and so are not amenable for simulating quantum magnetism in the fashion discussed in this work.}.  The basic idea is to encode effective degrees of freedom in long-lived, low-lying, and easily accessible internal degrees of freedom such as rotational and vibrational modes.  Dipole-dipole interactions  can mediate   coupling between these effective spins even when molecules are pinned  in a lattice (i.e.~when their tunneling is completely suppressed).  One consequence is that  polar molecules can be  used to study far-from-equilibrium quantum magnetism even in non-degenerate quantum gases.  Such far-from-equilibrium dynamics has been observed in ultracold polar molecule experiments and will play a central role in our exposition.  Additionally, taking a broad view of molecular diversity and experimental constraints such as temperature and lattice-scale probe resolution, we  provide an overview of the "quantum simulation landscape" of quantum magnetic phenomena achievable with ultracold molecules, summarized in Fig.~\ref{fig:complexity}. This figure shows the new regimes of quantum magnetism that become available both as the motional entropy and temperature decrease (vertical axis) and as the complexity of the internal molecular structure increases (horizontal axis).  This figure will be discussed in more detail in Sec.~\ref{sec:PartIII}.

Our paper is organized as follows.  Sec.~\ref{sec:PartII} derives effective spin Hamiltonians describing the internal state dynamics of polar molecules in optical lattices (Sec.~\ref{sec:QMMol}) and neutral atoms in optical lattices for comparison (Sec.~\ref{sec:superexchange}), focusing on the simplest scenarios.  Sec.~\ref{sec:Dynamical} describes recent experiments in which effective quantum magnetism has been experimentally probed via far-from-equilibrium dynamics, as well as new theoretical tools that were developed to verify and understand the experimental observations.  In Sec.~\ref{sec:somethingaboutmolecules} we explore molecules with  complex internal structure, review methods of producing ultracold molecules, and go over basic molecular structure.  In Sec.~\ref{sec:PartIII} we identify future directions for quantum magnetism with ultracold molecules, considering both advances in experimental technology and the structural complexity of molecules on the experimental horizon.  Finally, in Sec.~\ref{sec:Concl}, we conclude.

\section{Quantum magnetism with ultracold molecules and atoms}
\label{sec:PartII}

Exploring quantum magnetism with ultracold matter is a particularly fruitful direction of research, as quantum magnetic phenomena lie at the core of condensed matter physics.  Moreover, despite the apparent simplicity of many models of quantum magnetism, these models are in general extraordinarily hard to solve with classical resources.  This makes them excellent candidates for ``quantum simulation" with ultracold systems.  Although quantum magnetism is a vast field that is well beyond the capacity of this review to cover, we mention here some of the broad ideas that motivate its study. More complete reviews and introductions can be found, e.g., in Refs.~\citen{Auerbach,sachdev_quantum_2008,lacroix_introduction_2011}. 

One reason for the intense study of quantum magnetism is its relevance to materials and experimental phenomena -- for example, antiferromagnets, multiferroic materials, spin glasses, and spin nematics --  and the frequent proximity of quantum magnetism to unconventional superconductivity. Another motivation is the numerous \textit{exotic} phenomena that are theoretically predicted, including topologically ordered phases and (algebraic) spin liquids. These harbor physics which cannot be described within the ``Landau paradigm" of symmetry breaking, as is also the case with the fractional quantum Hall effect.  Observing a broader range of such phenomena which lie outside of conventional classification would clearly deepen our knowledge of quantum many-body physics.  Finally, an understanding of quantum magnetism can have a great impact in advancing current technology including better and faster hard drives, computers  and spintronic devices.

Quantum magnetism in the solid state usually refers to interactions between electron spins localized in a crystal lattice.  As the Coulomb interaction, which provides the microscopic interaction between electrons, is spin-independent, interactions between spins should be interpreted as \emph{effective} interactions which arise from Coulomb interactions in conjunction with Fermi statistics, i.e.~the required antisymmetry of electrons under exchange.  In Sec.~\ref{sec:superexchange}, we show how such effective magnetic interactions arise from particles with short-range interactions when they can tunnel in a lattice. This spin interaction mechanism, known as superexchange, is  the most common mechanism by which effective magnetic phenomena arise in cold \textit{atomic} gases loaded in optical lattices.  Before we discuss the superexchange mechanism, however, Sec.~\ref{sec:QMMol} describes the simplest example of effective quantum magnetism mediated by dipole-dipole interactions in polar molecules, the main focus of this review.  As we shall see, the resulting models for dipole-mediated quantum magnetism and the superexchange mechanism are very similar, even though the physical mechanism is very different.  Finally, Sec.~\ref{sec:XXZ} discusses control and experimental consequences of the terms appearing in the effective spin models.

\subsection{Effective magnetism with polar molecules}
\label{sec:QMMol}

\begin{figure}[bp]
\centerline{\includegraphics[width=0.666\columnwidth]{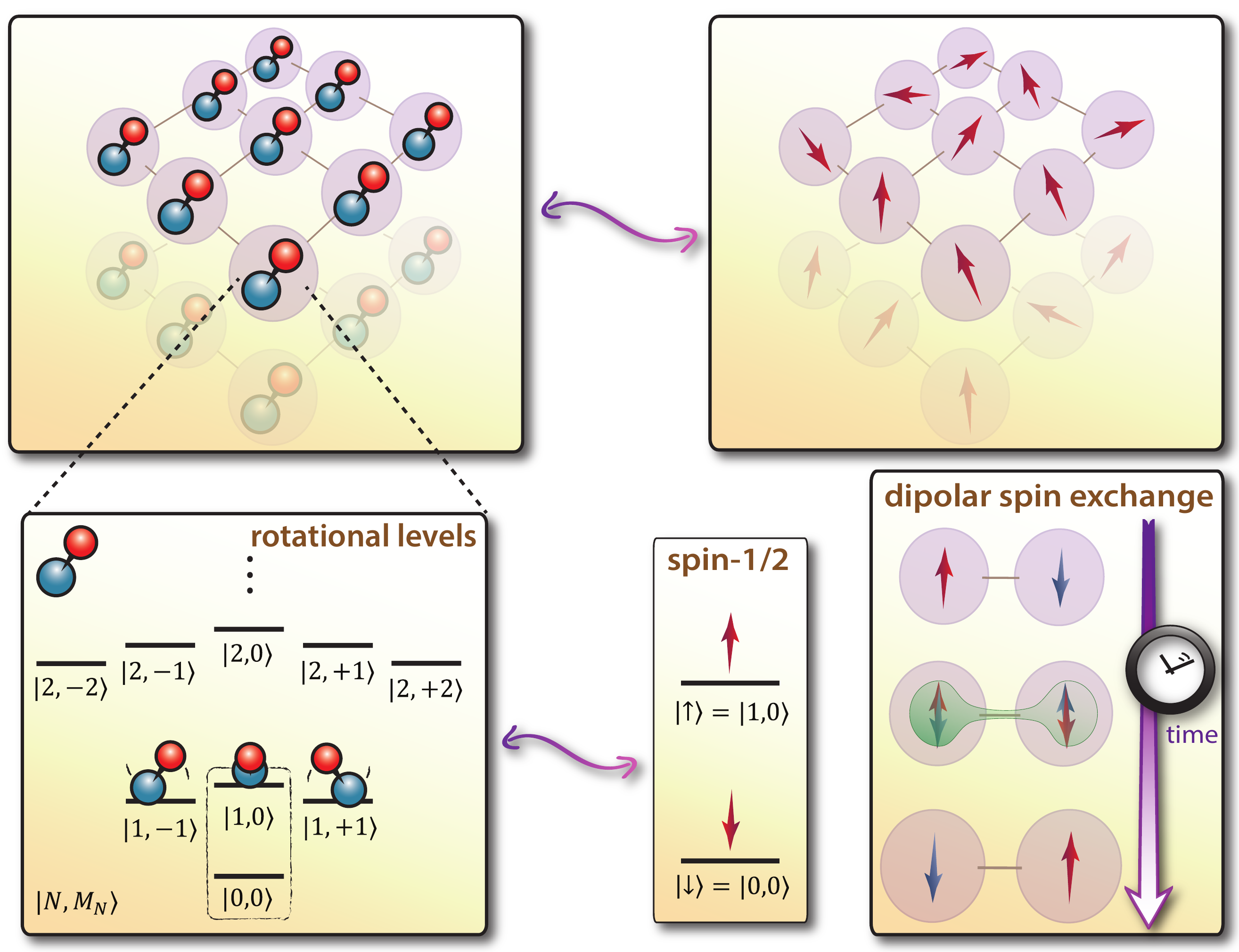}}
\caption{\label{fig:SpinModels}  Quantum magnetism of ultracold molecules in a lattice. 
Molecules in a deep lattice (top left) realize lattice spin models (top right), when the lattice is deep enough to \textit{suppress} tunneling.  The spin degree of freedom is encoded in rotational states of the molecule (bottom left).  Two types of interactions occur naturally: a "spin exchange" interaction that exchanges pairs of spin states (illustrated, bottom right) and, in the presence of a dc electric field, an "Ising" interaction that splits the energies of aligned and anti-aligned pairs of spins [see Eq.~\eqref{eq:dipolarXXZ}]. Both processes are capable of  correlating and entangling spins.
}

\end{figure}

Let us now consider how effective quantum magnetism arises for polar molecules~\cite{barnett:quantum_2006,Wall_PRA_2010,Gorshkov_Manmana_11,Gorshkov_Manmana_11b}.  For clarity, we discuss the simplest manifestation in this section before discussing how more complex magnetic interactions may be engineered in Secs.~\ref{sec:MolecularStructure} and~\ref{sec:PartIII}.  Our starting point is shown schematically in the top left panel of Fig.~\ref{fig:SpinModels}.  Here, molecules are pinned in a deep optical lattice with exactly one molecule per lattice site.  By "pinned" we mean that molecules do not move between lattice sites on the timescales of an experiment.  We now wish to encode an effective spin-1/2 in the internal degrees of freedom of the molecule.  Considering $^1\Sigma$ molecules, in which there are no unpaired spins or orbital angular momentum\footnote{A review of molecular structure and terminology is given in Sec.~\ref{sec:MolecularStructure}.}, and neglecting hyperfine structure, the lowest-lying degrees of freedom to encode spin in are the rotational degrees of freedom\footnote{In fact, these non-rotational degrees of freedom can sometimes be neglected in  molecules with more complex molecular structures under appropriate circumstances.}. The rotational states are described by a rigid rotor Hamiltonian~\cite{Zare} and can be labeled by $|NM_N\rangle $, where $N$ is the rotational angular momentum quantum number and $-N\le M_N\le N$ is the projection of the rotational angular momentum along a space-fixed quantization axis.  In the absence of external fields, the rotational energy spectrum is $E_{NM_N}=B_N N\left(N+1\right)$, where $B_N$ is called the rotational constant and is proportional to the inverse moment of inertia of the molecule.  Typical rotational constants are a few GHz, which is much larger than the dipolar interaction energies of molecules at typical $\sim 500$nm optical lattice spacings, and also much larger than ultracold temperatures.  These facts, together with the anharmonic spectrum and very long ($>10$s) lifetimes of rotational excitations, imply that the number of molecules in excited rotational states is conserved over the timescale of an experiment\footnote{Provided, of course, that rotational excitations are not generated by external means, e.g.~by a microwave field.}.

The $(2N+1)-$fold degeneracy of rotational excited states makes isolating a pair of rotational levels in which to encode a spin-1/2 challenging, and so we would like to split this degeneracy.  One way to do so is to introduce a DC electric field $\mathbf{E}_{\mathrm{DC}}$; the resulting splitting is illustrated in Fig.~\ref{fig:SpinModels}.   The key feature that we need in order to understand the emergence of quantum magnetism is that states such as  $|0,0\rangle$ and $|1,0\rangle$ in Fig.~\ref{fig:SpinModels}, are now well-isolated from all other states, and so we can use them to encode a spin-1/2. The details of the coupling and notation will be explained below.

Now that we have isolated an effective spin-1/2, we investigate the effect of dipole-dipole interactions within this subspace of states.  The dipole-dipole interaction between molecules $i$ and $j$ is
\begin{align}
\label{eq:DDI}\hat{H}_{\mathrm{DDI}}&=\frac{\hat{\mathbf{d}}_{i}  \cdot \hat{\mathbf{d}}_j-3\left(\hat{\mathbf{d}}_i\cdot \mathbf{e}_r\right)\left(\hat{\mathbf{d}_j}\cdot \mathbf{e}_r\right)}{r^3}\, ,
\end{align}
where $\mathbf{e}_r$ is a unit vector connecting  molecules $i$ and $j$, $r$ is the distance between these molecules, and $\hat{\mathbf{d}}_i$ is the dipole operator of molecule $i$.  
Within the subspace of states $\left\{|\downarrow\rangle,|\uparrow\rangle\right\}\equiv \left\{|0,0\rangle, |1,0\rangle\right\}$ forming our spin-1/2, and in the limit of the interaction being much smaller than the rotational splitting, the interaction Eq.~\eqref{eq:DDI} is simple; for example
\begin{equation}
\label{eq:DExam}\hat{H}_{\mathrm{DDI}} \ket{\uparrow \downarrow} =  a\ket{\uparrow \downarrow}+  b\ket{\downarrow\uparrow }
\end{equation} 
with $a=\bra{\uparrow \downarrow}\hat{H}_{\mathrm{DDI}}\ket{\uparrow \downarrow}$ and $b=
 \bra{\downarrow \uparrow}\hat{H}_{\mathrm{DDI}}\ket{\uparrow \downarrow} $.
Processes that change the total magnetization, such as $\ket{\uparrow \downarrow} \rightarrow \ket{\uparrow \uparrow} $,
are energetically far off-resonant and therefore negligible.
This implies that the spin-spin interaction of the molecules will be 
\begin{align}
\label{eq:HijSpin-prelim}\hat{H}_{ij}&= \left[\frac{J_{\perp}(\hat{\mathbf{d}}_i,\hat{\mathbf{d}}_j,\mathbf{r}_i-\mathbf{r}_j)}{2}\left(\hat{S}^+_i\hat{S}^-_j+\mathrm{h.c.}\right)+J_z(\hat{\mathbf{d}}_i,\hat{\mathbf{d}}_j,\mathbf{r}_i-\mathbf{r}_j)\hat{S}^z_i\hat{S}^z_j\right]\,
\end{align}
with $S^{\pm,z}_i$ the usual spin-1/2 operators acting on molecule $i$. 
We will derive below the forms of $J_\perp(\hat{\mathbf{d}}_i,\hat{\mathbf{d}}_j,\mathbf{r}_i-\mathbf{r}_j)$ and $J_z(\hat{\mathbf{d}}_i,\hat{\mathbf{d}}_j,\mathbf{r}_i-\mathbf{r}_j)$ and  their dependence on electric field and choice of rotational states, as well as determine additional single spin terms that are omitted in Eq.~\ref{eq:HijSpin-prelim} [see Eq.~\eqref{eq:HijSpin} for the final result]. 

However, before giving a more complete derivation of Eq.~\eqref{eq:HijSpin-prelim} and determination of the  coefficients in it, we first will describe how an electric field may be used to achieve the level splitting required to energetically isolate the spin-1/2 degree of freedom. This also provides useful background for how the interactions in Eq.~\eqref{eq:HijSpin-prelim} may be manipulated with electric fields.

 The coupling Hamiltonian of the molecule's dipole operator, $\hat{\mathbf{d}}$,  with the external field is $-\mathbf{d}\cdot \mathbf{E}_{\mathrm DC}$. We take $\mathbf{E}_{\mathrm DC}$ to set the space-fixed $z$ axis, $\mathbf{E}_{\mathrm{DC}}=E_{\mathrm{DC}}\mathbf{e}_z$ and thus the coupling Hamiltonian has matrix elements
\begin{align}
 \langle N'M_N'|-\hat{d}_zE_{\mathrm{DC}}|NM_N\rangle&=-E_{\mathrm{DC}}\delta_{M_N',M_N}\langle N'M_N'|\hat{d}_0|NM_N\rangle
\end{align}
where
\begin{align}
\label{eq:dipMEs}\langle N'M_N'|\hat{d}_p|NM_N\rangle&= d\left(-1\right)^{M_N}\sqrt{\left(2N'+1\right)\left(2N+1\right)} \left(\begin{array}{ccc} N&1&N\\ -M_N&p&M_N\end{array}\right)\left(\begin{array}{ccc} N'&1&N\\ 0&0&0\end{array}\right)\, 
\end{align}
are the matrix elements of the dipole operator in the basis of rotational states.  In Eq.~\eqref{eq:dipMEs}, $\left(\dots\right)$ is a 3$j$-symbol and $\hat{d}_{\pm 1}=\mp(\hat{d}_x\pm i \hat{d}_y)/\sqrt{2}$ and $\hat{d}_0=\hat{d}_z$ are spherical components of the dipole operator. Only the $p=0$ component is used above, but the $p=\pm1$ components will be useful to us later. The electric field mixes rotational states while preserving their projection $M_N$ on the field axis.  Because the electric field does not cause level crossings, we can still label the eigenstates of rotation in the presence of a DC field with $|NM_N\rangle$, where $N$ is now interpreted as a label corresponding to the number of rotational quanta if the field were to be ramped adiabatically  to zero.  The energies of these states in weak fields, calculated to lowest order in the small parameter $\beta_{\mathrm{DC}}=dE_{\mathrm{DC}}/B_N$, are
\begin{align}
E_{NM_N}/B_N&=N\left(N+1\right)+\frac{\beta_{\mathrm{DC}}^2}{2}\frac{N\left(N+1\right)-3M_N^2}{\left(2N-1\right)\left(2N+3\right)N\left(N+1\right)}\, .
\end{align}
As shown in Fig.~\ref{fig:SpinModels}, all states with the same value of $|M_N|$ remain degenerate, and are separated from all other states with the same value of $N$ by an energy $\sim \beta_{\mathrm{DC}}^2B_N$.

\begin{figure}[bp]
\centerline{\includegraphics[width=0.444\columnwidth]{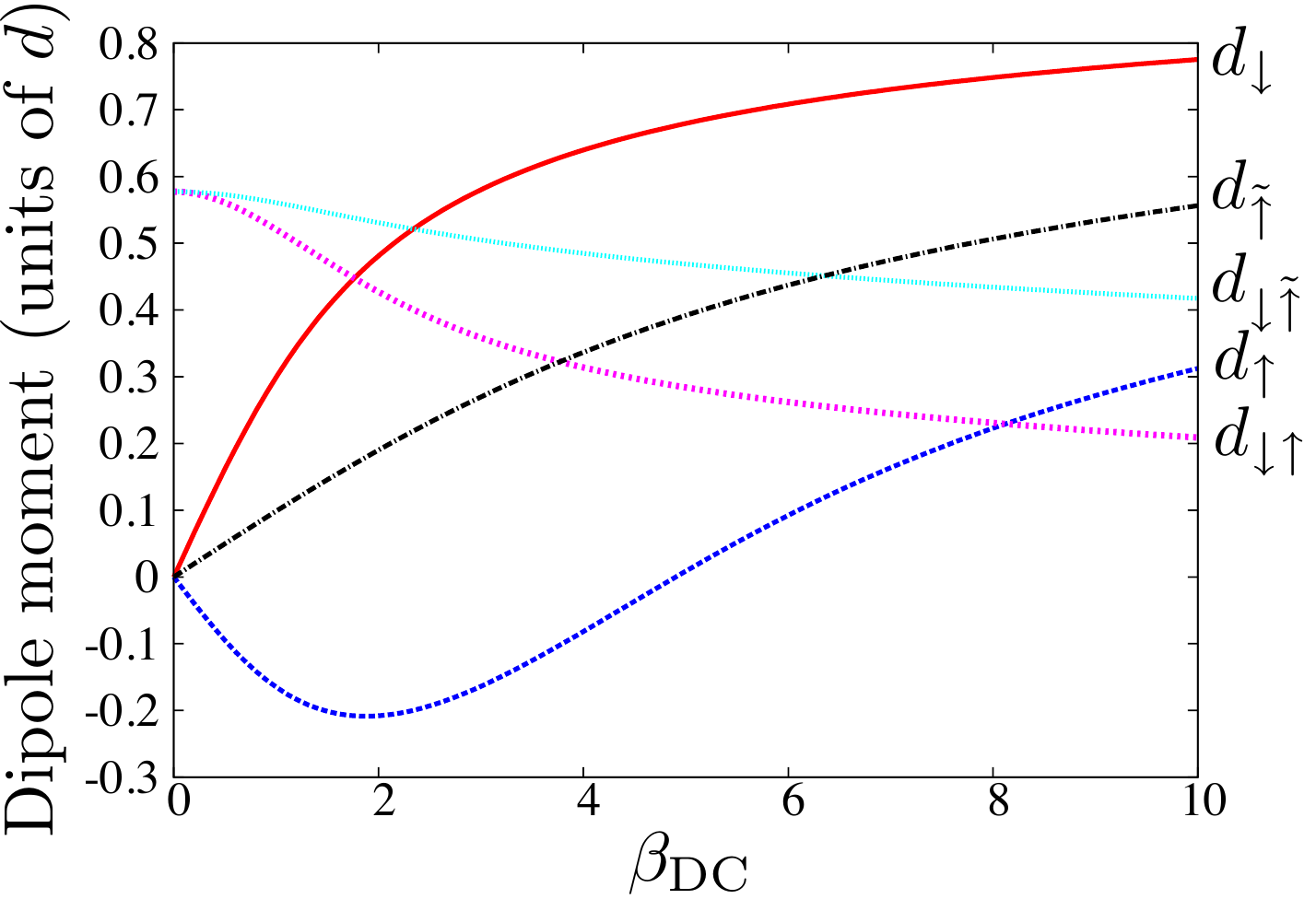}} 
\caption{   \label{fig:Efield}  
Tuning spin-spin interactions with an electric field.  An applied electric field with dimensionless strength $\beta_{\mathrm{DC}}=dE_{\mathrm{DC}}/B_N$ alters the expected dipole moments of the states $|\downarrow\rangle$, $|\uparrow\rangle$, and $|\tilde{\uparrow}\rangle$, see Eqs.~\eqref{eq:res}-\eqref{eq:trans} and \eqref{eq:res1}-\eqref{eq:trans1}.  In particular, the resonant dipole moments increase linearly for small fields and saturate asymptotically to the permanent dipole moment.  The transition dipole moments are nonzero in zero field and monotonically decrease as the field is increased.  The dipole moments determine the spin-spin couplings through, e.g., Eqs.~\eqref{eq:Jp1}-\eqref{eq:V1}.
}
\end{figure}

Now turning to the full derivation of how the dipole-dipole interaction, Eq.~\eqref{eq:DDI}, projects onto the spin-1/2 degree of freedom we note that the dipolar interaction Eq.~\eqref{eq:DDI} is the contraction of two rank-two tensors, one acting on the internal state of the molecules through the dipole operators and one acting on the orbital motion of the two molecules through the angular dependence of Eq.~\eqref{eq:DDI}.  Hence, the dipole-dipole interaction can be written as a sum of terms which transfer $q$ units of rotational angular momentum projection to orbital angular momentum projection with $-2\le q\le 2$.  Since our spin-1/2 is comprised of states with $M_N=0$, there is no way to change the rotational projection quantum number within this set of states (as is implicit in Eq.~\eqref{eq:DExam}).  Hence, only the $q=0$ terms are important.  Explicitly, these terms are
\begin{align}
\label{eq:DDI0}\hat{H}_{\mathrm{DDI}; q=0}&=\frac{1-3\cos^2\theta}{r^3}\left[\hat{d}_0\hat{d}_0+\frac{\hat{d}_1\hat{d}_{-1}+\hat{d}_{-1}\hat{d}_1}{2}\right]\, 
\end{align}
with $\theta$ the angle between the quantization axis and the vector connecting the two molecules.
Within the subspace of states $\left\{|\downarrow\rangle,|\uparrow\rangle\right\}= \left\{|0,0\rangle, |1,0\rangle\right\}$ the expectations of $\hat{d}_{\pm 1}$ all vanish due to selection rules.  The matrix elements of the $\hat{d}_0$ operator, computed using Eq.~\eqref{eq:dipMEs} to lowest order in $\beta_{\mathrm{DC}}$, are
\begin{align}
\label{eq:res}\langle \downarrow|\hat{d}_0|\downarrow\rangle &=\frac{\beta_{\mathrm{DC}}}{3}\equiv d_{\downarrow}\, ,\;\; \langle \uparrow |\hat{d}_0|\uparrow \rangle =-\frac{\beta_{\mathrm{DC}}}{5}\equiv d_{\uparrow}\, ,\\
\label{eq:trans}\langle \uparrow |\hat{d}_0|\downarrow\rangle&=\langle \downarrow |\hat{d}_0|\uparrow\rangle=\frac{1}{\sqrt{3}}\left(1-\frac{43 \beta_{\mathrm{DC}}^2}{360}\right)\equiv d_{\downarrow\uparrow}\, .
\end{align}
The \emph{resonant} dipole moments in Eq.~\eqref{eq:res} describe the degree of orientation of the states $|\downarrow\rangle$ and $|\uparrow\rangle$ in the presence of the external field.  Importantly, these resonant dipole moments vanish in zero field, which says that the average dipole moment of a pure rotational state along any direction fixed in space is zero.  In contrast, the \emph{transition} dipole moments in Eq.~\eqref{eq:trans} are nonzero even in zero field.\footnote{Physically, the dipole operator is a vector operator, and so is odd under parity.  The parity of the state $|NM_N\rangle$ is $(-1)^N$, and so nonzero matrix elements in the absence of an electric field can exist only between states $N$ and $N\pm1$.} These encapsulate the strength of a dipole-allowed transition from $|\downarrow\rangle$ to $|\uparrow\rangle$ and vice-versa.  For example, these transition dipole matrix elements determine the Rabi frequency of the transition from $|\downarrow\rangle$ to $|\uparrow\rangle$ when a molecule is illuminated with near-resonant microwave radiation.  Fig.~\ref{fig:Efield} shows the behavior of these dipole moments, going beyond the perturbative $\beta_{\mathrm{DC}}\ll 1$ limit.

Let us now consider two molecules at lattice sites $i$ and $j$, and write the dipole-dipole interaction, Eq.~\eqref{eq:DDI0}, in the basis $\{|\uparrow_i\uparrow_j\rangle, |\uparrow_i\downarrow_j\rangle,|\downarrow_i\uparrow_j\rangle,|\downarrow_i\downarrow_j\rangle\}$.  We find
\begin{align}
\label{eq:Hij}\hat{H}_{ij}&=\frac{1-3\cos^2\theta_{ij}}{\left|\mathbf{r}_i-\mathbf{r}_j\right|^3}\left(\begin{array}{cccc} d_{\uparrow}^2&0&0&0\\ 0&d_{\downarrow}d_{\uparrow}&d_{\downarrow\uparrow}^2&0 \\ 0&d_{\downarrow\uparrow}^2&d_{\downarrow}d_{\uparrow}&0\\ 0&0&0&d_{\downarrow}^2\end{array}\right)\,,
\end{align}
with $\mathbf{r}_i$ the location of molecule $i$ measured in units of the lattice spacing $a$.
If we define the spin-1/2 operators $\hat{S}^z_i=\frac{1}{2}(|\uparrow_i\rangle\langle \uparrow_i|-|\downarrow_i\rangle\langle \downarrow_i|)$, $\hat{S}^+_i=|\uparrow_i\rangle\langle \downarrow_i|$, $\hat{S}^-_i=(\hat{S}^{+}_i)^{\dagger}$ obeying the commutation relations\footnote{\label{ftnt:sphericaltens}Note that $\hat{S}^{\pm}=(\hat{S}^x\pm i \hat{S}^y)$ is the "raising operator" form of the spin operators, but the dipole operators $\hat{d}^{\pm 1}=\mp(\hat{d}^x\pm i\hat{d}^y)/\sqrt{2}$ are written in the standard form for spherical tensor operators (with the overall minus sign on $\hat{d}^{1}$).}  $\left[\hat{S}^z_i,\hat{S}^{\pm}_j\right]=\pm \delta_{ij}\hat{S}^{\pm}_i$, we can re-write Eq.~\eqref{eq:Hij} as a spin Hamiltonian:
\begin{align}
\label{eq:HijSpin}\hat{H}_{ij}&=\frac{1-3\cos^2\theta_{ij}}{\left|\mathbf{r}_i-\mathbf{r}_j\right|^3}\left[\frac{J_{\perp}}{2}\left(\hat{S}^+_i\hat{S}^-_j+\mathrm{h.c.}\right)+J_z\hat{S}^z_i\hat{S}^z_j+W\left(\mathbb{I}_i\hat{S}^z_j+\hat{S}^z_i\mathbb{I}_j\right)+V\mathbb{I}_i\mathbb{I}_j\right]\, ,
\end{align}
where 
\begin{align}
\label{eq:Jp1}J_{\perp}&\equiv 2d_{\downarrow\uparrow}^2\, ,\\
J_z&\equiv (d_{\uparrow}-d_{\downarrow})^2\, ,\\
W&\equiv (d_{\uparrow}^2-d_{\downarrow}^2)/2\, ,\\
\label{eq:V1}V&\equiv (d_{\uparrow}+d_{\downarrow})^2/4\, ,
\end{align}
and $\mathbb{I}$ is the identity operator in spin space. This result gives the full determination of the Hamiltonian motivated in Eq.~\eqref{eq:HijSpin-prelim}.   Eq.~\eqref{eq:HijSpin} readily generalizes to a many-site lattice in which each site can be filled with $n_i=0$ or $1$ molecules: one sums over all pairs $i$ and $j$, replaces the identity operator in spin space with the total molecular density operator $\mathbb{I}_i\to \hat{n}_i\equiv\sum_{\sigma=\uparrow,\downarrow}\hat{a}^{\dagger}_{i\sigma}\hat{a}_{i\sigma}$, and replaces the spin operators by their second-quantized counterparts 
\begin{align}
\label{eq:SQSpins}\hat{S}_i^{a}&=\frac{1}{2}\sum_{\mu \nu}\hat{a}^{\dagger}_{i\mu}\left[\sigma^a\right]_{\mu \nu}\hat{a}_{i\nu}\, .
\end{align}
Here, $\hat{a}_{i\nu}$ is a fermionic or bosonic operator in second quantization destroying a particle at site $i$ in spin state $\nu$ and $\sigma^a$ is a Pauli matrix. The term proportional to $V$ represents coupling of the molecule density to itself, and so is a constant for pinned molecules and may be ignored.  Similarly, in a unit-filled lattice the term proportional to $W$ is a constant of the motion (to the extent the density is homogeneous) and may be ignored.  The resulting effective spin model description for pinned molecules in a unit-filled lattice reads
\begin{align}
\label{eq:dipolarXXZ}\hat{H}&=\frac{1}{2}\sum_{i\ne j}\frac{1-3\cos^2\theta_{ij}}{\left|\mathbf{r}_i-\mathbf{r}_j\right|^3}\left[\frac{J_{\perp}}{2}\left(\hat{S}^+_i\hat{S}^-_j+\mathrm{h.c.}\right)+J_z\hat{S}^z_i\hat{S}^z_j\right]\,.
\end{align}
The Hamiltonian $\hat{H}$ is called an XXZ spin model, for reasons that we will clarify in Sec.~\ref{sec:XXZ}.  The effective magnetic interactions are long ranged, decaying as $1/r^3$ with the distance between lattice sites, and inherit the $(1-3\cos^2\theta)$ anisotropy characteristic of the dipole-dipole interaction, Eq.~\eqref{eq:DDI0}.  The relative importance of the $J_z$ and $J_{\perp}$ terms can be tuned by the strength of the external field, in accord with the $\beta_{\mathrm{DC}}$-dependence of the dipole matrix elements in Fig.~\ref{fig:Efield}.  In particular, the $J_z$ term (as well as the $W$ and $V$ terms, which can be neglected under some circumstances) vanishes in zero electric field.  We will investigate more complex means to tune effective magnetic interactions, even beyond the form in Eq.~\eqref{eq:dipolarXXZ}, in Sec.~\ref{sec:PartIII}.

The spin-spin interactions can also be modified by varying the choice of the two rotational levels used to encode the spin-1/2. To demonstrate this, let us perform the same analysis as above using the states $\{|\downarrow\rangle, |\tilde{\uparrow}\rangle\} \equiv \{|0,0\rangle,|1,1\rangle\}$.  The relevant dipole matrix elements, analogous to Eqs.~\eqref{eq:res}-\eqref{eq:trans}, are
\begin{align}
\label{eq:res1}\langle \downarrow|\hat{d}_0|\downarrow\rangle &=\frac{\beta_{\mathrm{DC}}}{3}\equiv d_{\downarrow}\, ,\;\; \langle \tilde{\uparrow} |\hat{d}_0|\tilde{\uparrow} \rangle =\frac{\beta_{\mathrm{DC}}}{10}\equiv d_{\tilde{\uparrow}}\, ,\\
\label{eq:trans1}\langle \tilde{\uparrow} |\hat{d}_1|\downarrow\rangle&=-\langle \downarrow |\hat{d}_{-1}|\tilde{\uparrow}\rangle=\frac{1}{\sqrt{3}}\left(1-\frac{49 \beta_{\mathrm{DC}}^2}{1440}\right)\equiv d_{\downarrow\tilde{\uparrow}}\, ,
\end{align}
where we have again used Eq.~\eqref{eq:dipMEs}.  Two differences are  apparent when comparing to Eqs.~\eqref{eq:res}-\eqref{eq:trans}.  Namely, $d_{\tilde{\uparrow}}$ is positive while $d_{\uparrow}$ was negative in Eq.~\eqref{eq:res} and the transition dipole matrix elements $d_{\downarrow\tilde{\uparrow}}$ are due to the operators $\hat{d}_{\pm1}$ instead of $\hat{d}_0$.  Another consequence of this is that when we project the $q=0$ part of the dipole-dipole interaction, Eq.~\eqref{eq:DDI0}, into the space of our effective spin-1/2, both the $\propto \hat{d}_0\hat{d}_0$ and $\propto (\hat{d}_1\hat{d}_{-1}+\hat{d}_{-1}\hat{d}_{1})$ terms contribute\footnote{For molecules with no nuclear spin structure, the $|1,1\rangle$ and $|1,-1\rangle$ states are degenerate and hence will be resonantly coupled by the $q=\pm2$ terms of the dipole-dipole interaction that were ignored when simplifying Eq.~\eqref{eq:DDI} to Eq.~\eqref{eq:DDI0}.  However, in the alkali dimers, e.g.~KRb, hyperfine structure splits this degeneracy by an amount large compared to typical interaction energies.  Hence, we do not consider the $q=\pm2$ terms here.}.  Projecting into the basis $\{|\tilde{\uparrow}_i\tilde{\uparrow}_j\rangle,|\tilde{\uparrow}_i\downarrow_j\rangle,|\downarrow_i\tilde{\uparrow}_j\rangle,|\downarrow_i\downarrow_j\rangle\}$, we find a matrix of the same form as Eq.~\eqref{eq:Hij}, only with $d_{\uparrow}$ replaced by $d_{\tilde{\uparrow}}$ and $d_{\downarrow\uparrow}^2$ replaced by $-d_{\downarrow\tilde{\uparrow}}^2/2$.  The factor of $1/2$ comes from the second term in the brackets of Eq.~\eqref{eq:DDI0}, and the minus sign from Eq.~\eqref{eq:trans1} (see also footnote \ref{ftnt:sphericaltens}).  The $-1/2$ can alternatively be semiclassically understood as coming from  the time average of two rotating dipoles.  Thus, the effective spin model description is again an XXZ model of precisely the same form as Eq.~\eqref{eq:dipolarXXZ}, the only difference being that $d_{\uparrow}\to d_{\tilde{\uparrow}}$ and $J_{\perp}=-d_{\downarrow\tilde{\uparrow}}^2$ rather than $2d_{\downarrow\uparrow}^2$.  When a magnetic coupling is negative, spins prefer to align to minimize the energy, resulting in a \emph{ferromagnetic} tendency.  Likewise, a positive value represents \emph{anti-ferromagnetic} coupling.  The two simple examples given above demonstrate that the form of the effective Hamiltonian, Eq.~\eqref{eq:dipolarXXZ}, namely the XXZ coupling and the $(1-3\cos^2\theta)/r^3$ dependence of interaction matrix elements, is a universal characteristic of the effective magnetic interactions between diatomic molecules prepared in two dipole-coupled rotational levels.  The tunable, non-universal aspects are the actual coupling constants $J_{\perp}$, $J_z$, etc.; these can be controlled and even made to change sign by external fields and the choice of rotational states forming the effective spin-1/2 (see, for example, Fig.~\ref{fig:Efield}).

\subsection{Effective magnetism with two-component atoms}
\label{sec:superexchange}

As an example of how quantum magnetism arises due to short-range, spin-independent interactions (in contrast to molecules), we consider the simplest model of interacting spin-1/2 fermionic particles on a lattice, the single-band Hubbard model
\begin{align}
\label{eq:Hubbard}\hat{H}_{\mathrm{Hubbard}}&=-t\sum_{\langle i,j\rangle}\sum_{\sigma\in \left\{\uparrow,\downarrow\right\}}\left[\hat{a}_{i\sigma}^{\dagger}\hat{a}_{j\sigma}+\mathrm{h.c.}\right]+U\sum_i\hat{n}_{i\uparrow}\hat{n}_{i \downarrow}\, .
\end{align}
Here, $t$ is the tunneling energy, $U$ the interaction energy, $\langle i,j\rangle$ represents a sum over all nearest-neighbor pairs of lattice sites $i$ and $j$, $\hat{a}_{i\sigma}$ is a fermionic operator in second quantization destroying a particle at site $i$ in spin state $\sigma$, and $\hat{n}_{i\sigma}=\hat{a}^{\dagger}_{i\sigma}\hat{a}_{i\sigma}$.  In a solid state system, $U$ represents an approximation to the screened Coulomb potential, and the Hubbard model is never expected to provide anything more than a crude guide to qualitative features.  However, ultracold neutral atoms dominantly interact via a short-ranged ($\sim 1/r^6$) van der Waals potential, which is well-modeled with a contact pseudopotential at low energies.  Consequently, the Hubbard model provides an excellent microscopic description of ultracold two-component fermionic atoms in an optical lattice~\cite{Jaksch_PRL_1998}.  In the ultracold atomic realization of the Hubbard model, two magnetic sublevels of the hyperfine manifold play the role of the electronic spin-1/2.

When $U\gg t$, an effective magnetic interaction arises between singly-occupied neighboring sites due to virtual occupation of sites by two fermions~\cite{Auerbach}, a mechanism known as \emph{superexchange}.  To see how this arises, let us consider two sites $i$ and $j$ which are nearest neighbors and are spanned by the basis $\{|\uparrow,\uparrow\rangle,|\uparrow \downarrow,0\rangle, |\uparrow,\downarrow\rangle, |\downarrow,\uparrow\rangle, |0,\uparrow \downarrow\rangle,|\downarrow,\downarrow\rangle \}$, where the left and right sides denote the occupations of sites $i$ and $j$, respectively, and the states $|\uparrow\downarrow,0\rangle$ and $|0,\uparrow\downarrow\rangle$ denote spin singlet states.  The Hubbard Hamiltonian in this basis is
\begin{align}
\hat{H}_{\mathrm{Hubbard}}&=\left(\begin{array}{cccccc} 
0&0&0&0&0&0\\
0&U&-t&-t&0&0\\
0&-t&0&0&-t&0\\
0&-t&0&0&-t&0\\
0&0&-t&-t&U&0\\
0&0&0&0&0&0 \end{array}\right)\, .
\end{align}
In the case $U\gg t$, tunneling to the energetically high-lying states $|\uparrow\downarrow,0\rangle$ and $|0,\uparrow\downarrow\rangle$ happens only "virtually," and so we can consider the effects of these states in lowest-order perturbation theory.  The effective Hamiltonian in the degenerate subspace spanned by $\{|\uparrow,\uparrow\rangle, |\uparrow,\downarrow\rangle, |\downarrow,\uparrow\rangle ,|\downarrow,\downarrow\rangle \}$, to lowest order in $t/U$, is
\begin{align}
\hat{H}_{\mathrm{eff}}&=\left(\begin{array}{cccc}
0&0&0&0\\
0&-2 t^2/U&-2t^2/U&0\\
0&-2t^2/U&-2t^2/U&0\\
0&0&0&0\end{array}\right)\, .
\end{align}
This effective Hamiltonian has a form similar to the matrix appearing in our derivation of effective magnetic interactions for polar molecules, Eq.~\eqref{eq:DDI}.  Accordingly, we can write this effective Hamiltonian as a spin model
\begin{align}
\label{eq:tJeff} \hat{H}_{\mathrm{eff}}&=\frac{J_{\perp}}{2}\left(\hat{S}^+_i\hat{S}^{-}_j+\mathrm{h.c.}\right)+J_z\hat{S}^z_i\hat{S}^z_j+V\mathbb{I}_i\mathbb{I}_j\, ,
\end{align}
where $J_z=J_{\perp}=4 t^2/U$ and $V=-t^2/U$.  Physically, the singlet state $(|\uparrow,\downarrow\rangle-|\downarrow,\uparrow\rangle)/\sqrt{2}$ can lower its energy by virtual tunneling to a doubly occupied site.  In contrast, the triplet states $|\uparrow,\uparrow\rangle$, $|\downarrow,\downarrow\rangle$, and $(|\uparrow,\downarrow\rangle+|\downarrow,\uparrow\rangle)/\sqrt{2}$ are not connected to any physical states by tunneling, as occupancy of a site by two particles of the same spin is forbidden by the Pauli principle.  Hence, quantum statistics plays a key role in the generation of effective magnetic couplings for spin-independent interactions.

Just as was the case for the polar molecules above, the two-site Hamiltonian, Eq.~\eqref{eq:tJeff}, can be immediately generalized to the many-site case.  Simplifications occur if we consider the case of half filling (one fermion per lattice site), since in this case tunneling processes involving only singly occupied sites do not occur, and the term proportional to $V$ becomes a constant of the motion which we neglect (again ignoring density inhomogeneities).  The resulting effective spin dynamics is then governed by the celebrated \emph{Heisenberg model}
\begin{equation}
\label{eq:HBM} \hat{H}_{\mathrm{Heisenberg}}=J\sum_{\langle i,j\rangle} \hat{\mathbf{S}}_i\cdot\hat{\mathbf{S}}_j\, ,
\end{equation}
where $J=4t^2/U$ and $\hat{\mathbf{S}}_i=(\hat{S}^x_i,\hat{S}^y_i,\hat{S}^z_i)$.  {The Heisenberg model, realized through the superexchange mechanism with a short-range interaction $U$, represents the minimal realization of quantum magnetism seen in ultracold gases in optical lattices.}  The spin coupling in the Heisenberg model is a special case of XXZ coupling with $J_{\perp}=J_z$.  An XXZ coupling can be realized in cold atom systems by superexchange with two-component bosons due to differences in interaction strengths between the two components~\cite{Duan}.  An important aspect of magnetism generated through superexchange is that it is mediated through motion via the tunnel-coupling $t$.  Hence, magnetic correlations are only visible when the temperature of the motional degrees of freedom is of order $J\sim t^2/U\sim$ 1$\,$nK.  This is an extremely cold temperature, even by ultracold  standards!

\subsection{Interpretation of processes in the XXZ model}
\label{sec:XXZ}
For spin$-1/2$s, the $J_{\perp}$ term in Eq.~\eqref{eq:dipolarXXZ} flips a spin up at site $j$ to spin down while simultaneously flipping a down spin at a neighboring site $i$ to up and vice versa.  Schematically, we can draw this process as $|\downarrow_i\uparrow_j\rangle\to |\uparrow_i\downarrow_j\rangle$, as in Fig.~\ref{fig:SpinModels}.  As this effectively swaps the spin labels of the two sites, we will refer to this process as \emph{spin exchange}\footnote{We remark that for higher-dimensional spin representations this term does not necessarily exchange the two spin components.  For example, two spin-1's can undergo the transition $|1,0\rangle |1,0\rangle\to |1,-1\rangle|1,1\rangle$ under the $J_{\perp}$ term.}.  One proposal to study this mechanism with polar molecules was in Ref.~\citen{PhysRevA.84.061605}.  Note that configurations of identical spin projections $|\uparrow_i\uparrow_j\rangle$ and $|\downarrow_i\downarrow_j\rangle$ do not participate in spin exchange.  The model Eq.~\eqref{eq:dipolarXXZ} with only spin exchange, $J_z=0$, can be written as
\begin{align}
\label{eq:XX}\hat{H}_{\mathrm{XX}}&=\frac{J_{\perp}}{4}\sum_{ i\ne j}\frac{1-3\cos^2\theta_{ij}}{\left|\mathbf{r}_i-\mathbf{r}_j\right|^3}\left(\hat{S}^+_i\hat{S}^-_j+\mathrm{h.c.}\right)\, ,\\
&=\frac{J_{\perp}}{2}\sum_{ i\ne j}\frac{1-3\cos^2\theta_{ij}}{\left|\mathbf{r}_i-\mathbf{r}_j\right|^3}\left(\hat{S}^x_i\hat{S}^x_j+\hat{S}^y_i\hat{S}^y_j\right)\, .
\end{align}
A spin Hamiltonian with such coupling is called an \emph{XX model}.  The nomenclature for an XXZ model, Eq.~\eqref{eq:dipolarXXZ}, should now be clear, as it represents equal coupling (XX) for the $x$ and $y$ components of the spins, with a different coupling $Z$ for the $z$ component of the spins.  In this same notation, Heisenberg coupling as in Eq.~\eqref{eq:HBM} would be called XXX coupling.

In the opposite limit $J_{\perp}\to 0$ of the XXZ model, we recover an \emph{Ising model}
\begin{equation}
\label{eq:Ising} \hat{H}_{\mathrm{Ising}}=\frac{J_z}{2}\sum_{ i\ne j}\frac{1-3\cos^2\theta_{ij}}{\left|\mathbf{r}_i-\mathbf{r}_j\right|^3}\hat{S}^z_i\hat{S}^z_j\, .
\end{equation}
  As opposed to the XX coupling, which causes transitions between pairs of spins which are anti-aligned in the $z$ basis and does not affect aligned spins, the Ising model affects all configurations and causes no transitions between configurations in the $z$ basis.  In particular, aligned configurations such as $|\uparrow_i\uparrow_i\rangle$ or $|\downarrow_i\downarrow_j\rangle$ have an Ising interaction energy $J_z$, while anti-aligned configurations $|\uparrow_i\downarrow_j\rangle$ or $|\downarrow_i\uparrow_j\rangle$ have an energy $-J_z$.  Because all operators appearing in the Ising model commute with one another, the static properties of the Ising model are effectively classical, and it displays no quantum phase transitions.  Nevertheless, the dynamics is highly \textit{nonclassical}, generating strong correlations and maximally entangled states~\cite{hazzard:quantum-correlations_2014}; however, because of its special commuting property, the dynamics of correlation functions in generalized Ising models, in which the $z$ components of spins interact on arbitrary lattices with arbitrary couplings, can be determined analytically~\cite{PhysRevA.87.042101,1367-2630-15-8-083007,1367-2630-15-11-113008}.

\section{Dynamical investigations of quantum magnetism: experiment and theory}
\label{sec:Dynamical}
 
In this section, we describe the experimental realization and observation of the dipolar XXZ model Eq.~\eqref{eq:dipolarXXZ}  in ultracold molecules and the theoretical techniques that were developed in order understand the resulting behavior.
Subsection~\ref{sec:expt-obs-spin-exchange} describes the experimental evidence that the ultracold polar molecule  experiments are operating in the regime quantitatively described by Eq.~\eqref{eq:dipolarXXZ}. This evidence is based on extensive measurements of far-from-equilibrium dynamics and comparisons with theory.

When the experiments were performed, their dynamics was beyond the ability of existing theoretical methods to treat quantitatively. As a result, they stimulated the 
development of new theoretical ideas and methods.  Subsection~\ref{sec:quantum-simulation-theory} describes techniques that have been developed with the aid of experiments and applied to them, and it gives a short survey of  theoretical  methods for which ultracold molecule experiments are expected to provide a  testing ground.

\subsection{Observation of spin-exchange and verification of Eq.~\eqref{eq:dipolarXXZ} \label{sec:expt-obs-spin-exchange}}

Reference~\citen{hazzard:far-from-equilibrium_2013} theoretically argued that the dipolar spin-spin interactions in Eq.~\eqref{eq:dipolarXXZ} could be observed  even in systems with lattice filling (number of particles divided by number of lattice sites) well below unity, extending  to densities so low that more sites are empty than are occupied. The proposal was to observe dynamics using Ramsey spectroscopy, a standard dynamic protocol in atomic and molecular experiments (described below). We will describe how three goals of this proposal have now been successfully  achieved: observation of spin-spin interactions in ultracold molecules,  benchmarking of the accuracy of Eq.~\eqref{eq:dipolarXXZ} to describe these interactions, and the exploration of physics beyond the regime accessible to previously existing exact methods.

\begin{figure}[bp]
\centerline{\includegraphics[width=0.666\columnwidth]{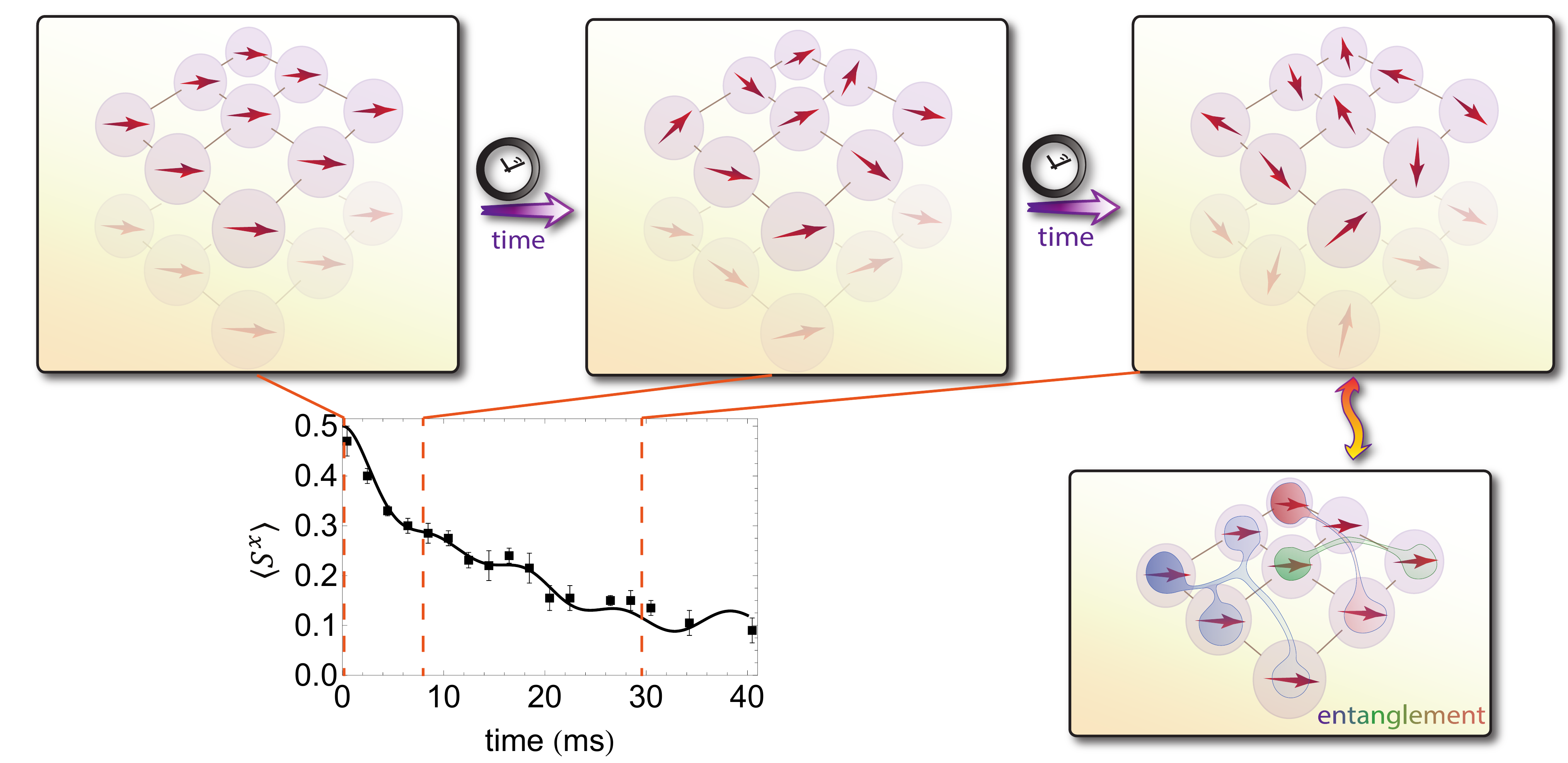}}
\caption{ \label{fig:dynamic-protocols} 
Non-equilibrium spin dynamics in a lattice. Bottom left plot and top dynamic sequence: initially aligned spins (along the $\hat x$ direction) evolve in time due to interaction.  Both inhomogeneous precession and the growth of entanglement can occur.  Such dynamics, implemented using Ramsey spectroscopy, has been used in ultracold KRb experiments to verify that the system accurately realizes Eq.~\eqref{eq:dipolarXXZ}. Related protocols can also probe the state of interesting physical systems. 
}

\end{figure}

Yan~\textit{et al.} employed Ramsey spectroscopy, illustrated in Fig.~\ref{fig:dynamic-protocols}, to observe spin-exchange interactions in ultracold KRb~\cite{yan:observation_2013}.    Ramsey spectroscopy may be described as follows (see also Fig.~\ref{fig:dynamic-protocols}): all of the spins are aligned along the same direction at time $t=0$,  then they are allowed to evolve dynamically under no external influence -- only Eq.~\eqref{eq:dipolarXXZ} -- for a time $t$, and finally one reads out the spin component along a chosen direction ${\hat n}$ summed over all particles, $ \expec{\sum_i \mathbf{S_i}\cdot {\hat n}}$.\footnote{
In the lab, the Ramsey protocol is implemented as a pair of microwave pulses, separated in time, that couple the rotational states.  The first aligns the spins in the desired direction, and the latter rotates the spins so that the various spin components can be measured by doing standard number measurements of each rotational state. Furthermore,  experiments so far have used a related, but  more robust measurement,  than choosing a single spin readout direction $\hat n$.  Instead, this vector is swept around a circle of constant $z$ component and the resulting ``Ramsey contrast" -- the oscillation amplitude as a function of this angle -- is the measured variable at each time. Moreover, the data we will show is taken with a spin-echo sequence, which adds an extra pulse to (partially) remove the effects of an inhomogeneous field $\sum_i h_i S^z_i$ that is present due to ``non-magic" trapping conditions~\cite{neyenhuis:anisotropic_2012}, i.e. the fact that trapping frequencies for the two rotational states differ slightly in this experiment.  
}

An example measurement of such dynamics, similar to Refs.~\citen{yan:observation_2013,hazzard:many-body_2014}, is shown in the graph at the bottom left of Fig.~\ref{fig:dynamic-protocols} for $N=11,000$ molecules in the lattice and with the spin-1/2 degree of freedom realized in the $\ket{\uparrow}=\ket{1,0}$ and  $\ket{\downarrow}=\ket{0,0}$  rotational states (see Fig.~\ref{fig:SpinModels}).  This panel shows the measured spin component along the $x$ direction as a function of the evolution time.  Experiments conclusively identify dipolar interactions from a number of these measurements; there are four  main pieces of evidence, which we recall now.

\begin{figure}[htp]
\centerline{\includegraphics[width=0.54\columnwidth]{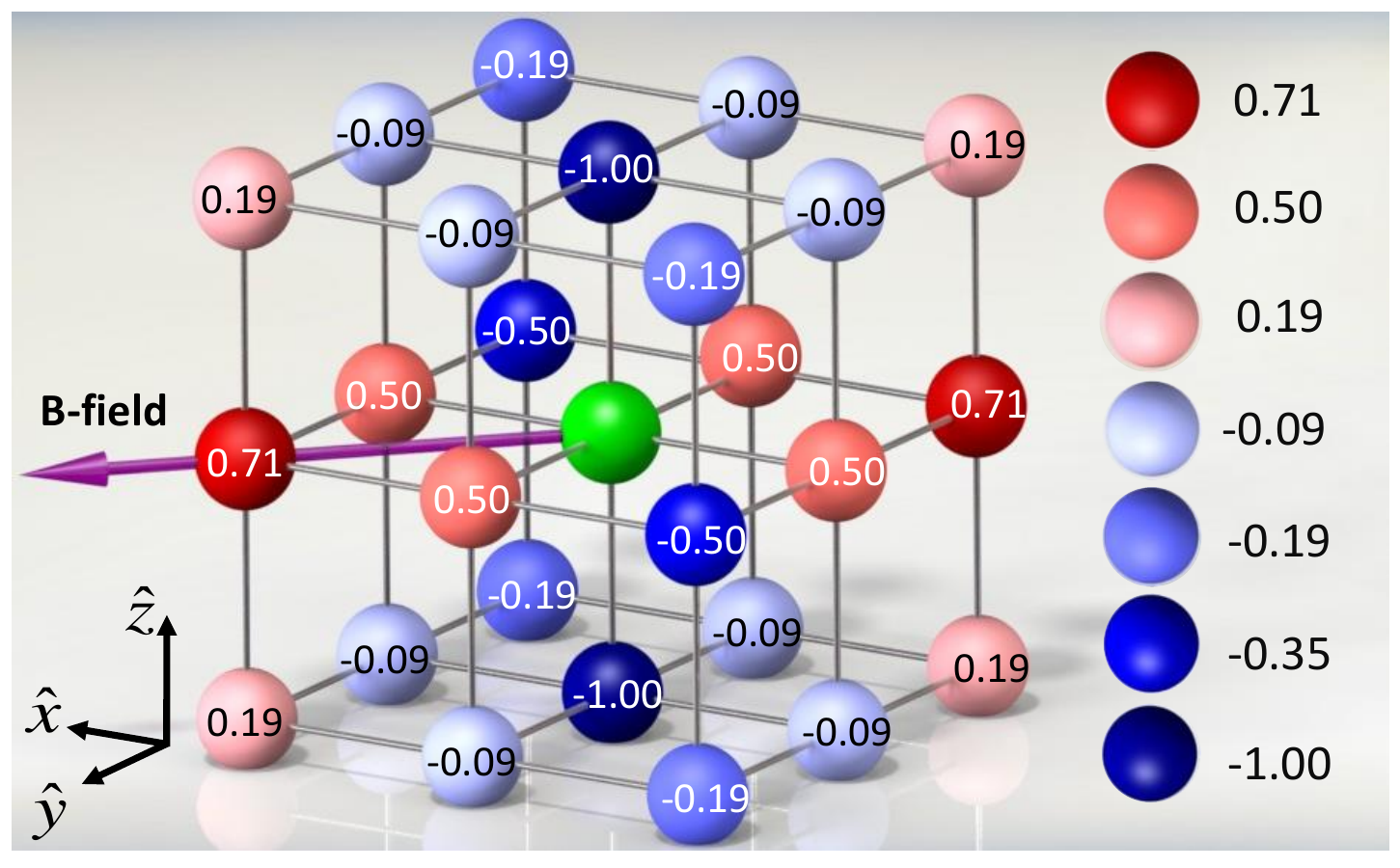}}
\caption{\label{fig:couplings-cube} Spin-spin interaction strengths. Labels on each site indicate the magnitude of the interaction between that site and the central green site, relative to the maximum interaction strength. This is given by the function $1-3\cos^2\theta$, where $\theta$ is the angle between  the intermolecular separation vector and the quantization axis. We take the quantization axis to be along the axis, halfway between the $\hat x$ and $\hat y$ directions, corresponding to the experimental set up of Ref.~\citen{yan:observation_2013}. [Here the \textit{magnetic} field $\mathbf{B}$, rather than an electric field, sets the quantization axis through nuclear magnetic effects.] Figure after Ref.~\citen{yan:observation_2013}.
}
\end{figure}

Two features are  apparent in Fig.~\ref{fig:dynamic-protocols} (bottom left): a relatively small  oscillation superposed on an exponential decay to zero.  Both of these features are due to dipolar spin-exchange interactions.  A first indication is that the observed oscillation frequency $\nu$ coincides  with the expected oscillation frequency of two KRb molecules in these rotational states occupying nearest neighbor sites separated along the $\hat z$ lattice direction (see Fig.~\ref{fig:couplings-cube}) namely $\nu=J_\perp/(2h)\approx 100$Hz.  The direction is relevant because of the $1-3\cos^2\theta$ anisotropy of the dipolar interaction, illustrated in Fig.~\ref{fig:couplings-cube}.  That this single coupling is so important is not an accident: the current experiments estimate about a $f=0.1$ lattice filling, and in this regime, the oscillations come from relatively rare configurations of molecules having a single nearest neighbor. These configurations have clear oscillatory dynamics, and the fastest frequency is the most visible~\cite{yan:observation_2013,hazzard:far-from-equilibrium_2013,hazzard:many-body_2014}.  Further data analysis was,  in fact, able to identify in the data other oscillation frequencies $\nu/\sqrt{2}$ and $\nu/2$, which corresponded to the next strongest interaction strengths~\cite{hazzard:many-body_2014}.

A second signature is that the time-constant for decay of $\expec{S^x}$ decreases with increasing density. In contrast, any single particle effect would remain unaffected by a simple change of density.   Further indicative of interactions, the coherence time was observed to decreased as $1/N$, with $N$ the total number of molecules. This scaling is   consistent with the
power law decay  of the dipolar interactions:
Increasing the lattice filling fraction $f\propto N$, decreases the mean distance
between molecules ($\bar{R}\propto f^{-1/3}$), leading to
an average dipolar interaction that scales as $1/\bar{R}^3\propto f\propto N$.

Two additional pieces of evidence were presented in Refs.~\citen{yan:observation_2013}  and~\citen{hazzard:many-body_2014} that we describe very briefly here.  First, the experiments in Ref.~\citen{yan:observation_2013} showed that some of the effects of interactions could be suppressed by using a more involved sequence of microwave pulses than the simple Ramsey spin-echo sequence, specifically the ``WAHUHA" sequence~\cite{waugh:approach_1968}.  This suppression was expected because this pulse sequence removes interactions completely in the case of two isolated molecules and is expected to partially remove the interactions in general.
Second, as discussed above, changing the rotational state pair used for the spin-1/2 system from $\{\ket{\uparrow}=\ket{1,0},\ket{\downarrow}=\ket{0,0}\}$ to $\{\ket{\uparrow}=\ket{1,-1},\ket{\downarrow}=\ket{0,0}\}$ decreases the interaction strength by a factor of two, but otherwise leaves the interaction unchanged. 
The experiments in Ref.~\citen{hazzard:many-body_2014} repeated the measurements with this new rotational state pair choice and showed that only the timescale, not the shape, of the $\expec{S^x(t)}$ dynamics changed when the interaction strength was varied.  This shows that \textit{only} the expected dipolar interactions contribute to the evolution over the measurement time.

These four pieces of evidence -- oscillation frequency, density-dependent contrast decay rate, response to the WAHUHA pulse sequence, and dependence of the dynamics on the rotational state used   -- clearly show that the JILA experiments have observed dipolar ``spin-exchange" interactions between molecules in different rotational states,  but on their own provide no direct evidence that the spin model describing the system is Eq.~\eqref{eq:dipolarXXZ} and do not tell what role various aspects of this equation play in the dynamics. For example, the evidence described above does not constrain whether the long-range interactions are playing a crucial role, and whether the spin interactions are of the XXZ form claimed.  The next subsection will describe how these questions were addressed in Ref.~\citen{hazzard:many-body_2014}. Doing this required comparing experiments to calculations using newly developed theoretical methods.
This subsection also will discuss other theoretical methods expected to be useful in the future.

We raise one broader point before turning to the theoretical comparison. All of the observations discussed were possible despite the non-degenerate temperature of the molecules in the lattice  and the consequent low filling. The key capability that enabled complex quantum many-body behavior to manifest in these measurements was the ability to perform precise quantum control of \textit{internal} degrees of freedom, namely the rotational states, i.e. the ability to prepare almost pure states in the spin degree of freedom.
Another aspect of these experiments is the decoupling of the spin and motional degrees of freedom: the motion is completely frozen in the lattice, and only affects the spin degrees of freedom by determining their (static) coupling strengths.  
It is interesting to note that the ability to control  the internal spin degrees of freedom has been utilized in other non-degenerate systems -- hydrogen gases~\cite{johnson:observation_1984}, optical lattice alkaline earth atomic clocks~\cite{lemke:p-wave_2011,martin:quantum_2013},  and warm (even room temperature)  alkali vapors in highly excited Rydberg states~\cite{stoneman:resonant-collision_1987}  -- as well as degenerate systems, including interacting spin-1/2 Fermi gases~\cite{demarco:spin_2002,koschorreck:universal_2013,bardon:transverse_2013}, and high-spin Fermi gases~\cite{krauser:giant_2014} (a highly non-comprehensive list). 
An intriguing question, therefore, is to what extent the quantum correlated physics survives at elevated temperatures in the presence of spin-motion decoupling. 
In many cases these prior experiments were not thought about as strongly interacting many-body systems  and it would be interesting to revisit them from this perspective with this question in mind.  One area that is particularly relevant to this review is the study of molecules at cold, but non-degenerate or possibly even not-ultracold temperatures. There are several experiments  aiming to produce or that have produced molecules in this regime (see Sec.~\ref{sec:MolecularStructure}).

\subsection{Quantum simulation of strongly correlated quantum magnetism: theoretical developments \label{sec:quantum-simulation-theory}}

 In order to quantitatively validate the molecules' realization of Eq.~\eqref{eq:dipolarXXZ}, it is necessary to compare their behavior to theoretical predictions. However, as discussed below, this required a new theoretical method, as all theories that existed prior to these experiments were extremely inaccurate for this problem. 
Here we focus on the method introduced in Ref.~\citen{hazzard:many-body_2014}, termed the MACE (moving average cluster expansion).

To understand how rich the physics of these systems is, as well as the complexity of calculating their dynamics, note that these systems combine several features that would seem to render the calculations intractable: they are three dimensional, long-ranged interacting, disordered, far-from-equilibrium quantum many-body systems. Moreover,  the dynamics is \textit{completely} beyond mean-field theory: mean-field theory  predicts \textit{no} dynamic evolution of the contrast for the initial conditions with all spins along the $\hat x$ direction, which is the situation  studied experimentally in Refs.~\citen{yan:observation_2013,hazzard:many-body_2014}. The absence of mean-field dynamics is most easily seen for the Ising term: the mean field Hamiltonian contains terms of the form $S^z_i\langle S^z_j\rangle$, and these expectations vanish if the spins point along $\hat x$.  A similar type of  analysis can be done  for the XX term, and this analysis shows that there is no mean field dynamical evolution due to this term, either.

Even in one dimension, where special theoretical techniques can be applied to solve for the dynamics,  
one finds that substantial entanglement emerges. The amount of entanglement generally increases even beyond that exhibited by the ground state~\cite{hazzard:far-from-equilibrium_2013}. Because of all of these features, the dynamics might  appear to be nearly impossible to solve exactly, either analytically or numerically. Indeed, the methods that existed prior to the experiments fail to capture it~\cite{hazzard:many-body_2014}.

In spite of these challenges, Ref.~\citen{hazzard:many-body_2014} developed a new  theoretical method -- the MACE -- to accurately calculate the $\expec{S^x(t)}$ dynamics of Eq.~\eqref{eq:dipolarXXZ} and applied it to the experiments. The idea of MACE is to break the system into independent clusters, but calculate observables (e.g. $\expec{S^x_i(t)}$) only for the spin $i$ at the center of the cluster. This avoids effects of the cluster boundary, at least until times where correlations have had time to propagate inward to the cluster center. Specifically, for each spin $i$ one creates a cluster of $g$ spins by enumerating  the $(g-1)$ largest coupling constants connecting $i$ to other spins $\{j\}$.  One exactly solves the dynamics for each cluster, and then calculates $\langle S^x_i(t)\rangle$ for each $i$ using the cluster optimized for spin $i$.
It turns out that the MACE  converges orders-of-magnitude faster than  previously existing state-of-the-art methods. The MACE calculations quantitatively agree with experiment with no fitting parameters, for a wide range of molecule lattice fillings and for different choices for rotational state pairs to realize the spin-1/2 degrees of freedom. (The filling fraction was determined from one dataset, and then this parameter was used for comparison with all of the numerous other datasets.) This agreement confirmed that Eq.~\eqref{eq:dipolarXXZ} accurately describes the molecules.  Moreover alternative models are ruled out, suggesting that potential experimental imperfections do not alter the interaction dynamics. For example, one clearly observes the long-range nature of the interactions in this manner, as truncating the theoretical calculation to nearest neighbor interactions fails to reproduce the dynamics even qualitatively.

Even for the MACE,  it is challenging to rigorously and quantitatively assess its convergence, and consequently the observed quantitative agreement with experimental measurements of $\expec{S^x(t)}$ also provides independent evidence that the MACE is converging. In this sense,  the experiments are serving as quantum simulators: systems whose dynamics are beyond the ability of existing methods to simulate in a controlled way. These experiments and newly developed theory are all the more interesting because the XXZ spin model is realized in numerous important systems, even outside of ultracold atomic and molecular gases, such as excitons in solids and  large molecules, nitrogen-vacancy (NV) centers in diamond, and magnetic defects  in other solid state systems.

We emphasize that the capability of the MACE to model the particular measurements carried out so far, however, does not mean that it will be able to capture other measurements -- even those accessible to current experiments. 
 Examples of experiments likely beyond the capacity of MACE  are measurements of spin transport and correlation functions, 
which involve connections between two or more regions that are several lattice sites apart.
In those cases,  the experiments themselves  are expected to shed light on
the dynamics dictated by  Eq.~\eqref{eq:dipolarXXZ} and will help  to benchmark theories used to model it.

Although at present it is unclear what the most promising theoretical methods will be for calculating correlations, transport, and other future experimental scenarios, we mention techniques that are likely to provide insight: 
\begin{itemize}
\item \textbf{Mean-field theory.} Although entirely inapplicable to the $\expec{S^x(t)}$ dynamics initiated from a state polarized along the $\hat x$ axis that we have concentrated on here, mean-field theory can be a useful tool in general. 
\item \textbf{MACE.} The MACE and further development of this method may capture much  behavior qualitatively, even in cases where  it ceases to be quantitatively accurate. 
\item \textbf{DMRG.} In  one dimensional systems, density matrix renormalization group (DMRG) methods~\cite{white:density_1992,vidal:efficient_2004,daley:time-dependent_2004,schollwoeck:density-matrix_2005,schollwoeck:density-matrix_2011}  can solve for a variety of behaviors in these spin models. It is possible to extend calculations to quasi-1D "ladders" of width $\sim 5$, to obtain insight into two-dimensional systems~\cite{doi:10.1146/annurev-conmatphys-020911-125018}.  
\item \textbf{Many-body perturbation theory.} In special  cases, perturbation theory   allows one to accurately calculate response properties. For example if a simple spin state is driven slightly out of equilibrium -- for example, the all-down spin state $\ket{\cdots \downarrow \downarrow \cdots}$ has a few spins flipped -- the spin excitations created are dilute and their behavior may be tractable with  many-body Green's function techniques (see, e.g., Refs.~\citen{abrikosov_methods_1975,negele_quantum_1998}).
\item \textbf{Truncated Wigner approximation.} The truncated Wigner approximation (TWA)~\cite{polkovnikov:phase_2010}  introduces the quantum noise in the initial state and then propagates the mean-field equations of motion. This is a promising route since in many of the proposed current experiments, the initial state has a product state structure and thus relatively small quantum noise. Consequently, the TWA may provide a way to capture many features of time  dynamics, out to some time that, while not too large, may go well beyond a naive short-time perturbation theory, and, in some cases, even the measurement capabilities of the experiment.
\item \textbf{Exact diagonalization.} For a system with a finite number of spins, in principle one can numerically construct the full Hamiltonian as a matrix in the finite Hilbert space, and then one can diagonalize this matrix using standard algorithms~\cite{noack:diagonalization_2005}.  However, the accessible system sizes are severely limited, because the Hilbert space dimension grows exponentially, like $2^N$ for an $N$ particle spin-1/2 system. In particular, studying more than $\sim 30$-$35$ spins is generally impossible.  Nevertheless, understanding the behavior for these relatively small systems can lend much insight into the general behavior. 
\end{itemize}
It will be fascinating to test the zoo of many-body approximations with experiments, discovering when various techniques are accurate, when they are not, and stimulated by this, discovering new concepts and methods.

\subsection{Future directions for molecules out of equilibrium}

As suggested above, there are many interesting facets that can be explored fairly immediately with the ongoing experiments. One general possibility is  measurement of spin conductivities in strongly interacting dipolar spin models. This includes the study of many-body localization~\cite{basko:many-body_2006,pal:many-body_2010,mathey:dynamics_2011,mathey:light_2010,yao:many-body_2013,kwasigroch:bose-einstein_2013,serbyn:interferometric_2014}, see Sec.~\ref{sec:PartIII}. Another possibility is to measure correlation growth~\cite{hazzard:quantum-correlations_2014} either by looking at global fluctuations in the Ramsey signal~\cite{foelling:quantum_2014,rey:probing_2014} (analogous to noise correlation spectroscopy~\cite{foelling:quantum_2014}) or by employing local spin flips and probes~\cite{knap:probing_2013}. 
 Another possibility is to explore intrinsically dynamic phenomena and look for dynamical phase transitions or universality out of equilibrium~\cite{Bistritzer:intrinsic_2007,degrandi:quench_2010,degrandi:quench-sine-gordon_2010,gritsev:universal_2010,mitra:time_2012,mitra:thermalization_2012,mitra:mode-coupling_2011,polkovnikov:nonequilibrium_2011,
 karrasch:luttinger-liquid_2012,vosk:many-body_2012,dallatorre:dynamics_2012,dallatorre:universal_2013,
sarkar:perturbative_2013,mitra:correlation_2013,acevedo:new_2014,hazzard:quantum-correlations_2014}.

\section{Increasing molecular complexity}
\label{sec:somethingaboutmolecules}

The simple derivation of an effective spin Hamiltonian for polar molecules in Sec.~\ref{sec:QMMol} assumed that the molecules had no internal structure besides the rotational degrees of freedom.  In Sec.~\ref{sec:PartIII} we will  describe what new types of magnetic phenomena occur for molecules with more complex structure.  The present section discusses progress towards and challenges for the production of ultracold molecules, focusing on molecules which are under active experimental investigation.  An excellent overview of all ultracold molecule experiments and their stage of progress circa 2012 may be found in Ref.~\citen{Quemener_Julienne_12}.  Sec.~\ref{sec:MolecularStructure} provides a brief review of molecular structure and terminology in order to keep our discussion self-contained and accessible to researchers outside the field of molecular physics.

\subsection{Cooling and trapping of molecular gases}

Laser cooling  is a workhorse of ultracold atomic physics. It removes entropy from a selectively excited atomic sample by utilizing spontaneous emission to return the system to a fiducial state.  Naturally, one wishes to extend this technology to molecules, and indeed the first proposal to do so appeared one year after the realization of atomic BEC~\cite{Stwalley}.  However, certain difficulties become immediately evident when attempting to laser cool molecules.  In particular, a molecule in an excited electronic state can spontaneously radiate ("branch") to one -- or many -- states that are in the ground electronic manifold, but are rotationally or vibrationally excited.  Hence, rather than returning to the ground state and lowering entropy, a molecule can distribute the photon energy among its various internal modes.  The vibrational modes are the most troublesome in this respect, as transitions between then do not obey strict selection rules, but are only governed by state overlaps known as Franck-Condon factors.  Simply adding additional lasers to pump molecular population out of unwanted levels back into the cooling cycle, as is done for hyperfine branching in atoms, adds additional complexity which quickly becomes unsustainable as molecular complexity itself increases.  Hence, the application of laser cooling to molecules has instead focused on special classes of molecules in which the vibrational spectrum is almost decoupled from the electronic structure, resulting in nearly closed vibrational cycling transitions~\cite{DiRosa}.

Since direct laser cooling of molecules using the standard techniques developed for atoms is generally difficult, the most successful ultracold polar molecule experiments to date instead create ultracold molecules by "assembling" atoms which have themselves been brought to quantum degeneracy.  For example, in the KRb experiment at JILA, this is performed by magneto-association of ultracold gases of K and Rb across a Feshbach resonance~\cite{Ni_Ospelkaus_08}\footnote{It is amusing to note that the first near-degenerate ground state molecule was formed by assembling the first bosonic (Rb) and fermionic (K) atoms to reach quantum degeneracy.}.  Many other such intra- and inter-species resonances have been identified.  The resulting molecules are very highly excited, but may be transferred to the ground rotational-vibrational-electronic state using stimulated Raman adiabatic passage (STIRAP)~\cite{Shapiro}, a feat which is enabled by modern developments in highly-stable laser technology.  Once in the rotational-vibrational-electronic ground state, molecules can be transferred to their hyperfine ground state by using further manipulation with microwaves~\cite{Ospelkaus_Ni_10}.  More details on the production of ultracold molecules can be found in a number of other recent reviews~\cite{Meerakker, Hutzler, Carr_Demille_09, KochandShapiro, Koehler}.

Polar molecules interact through electric dipole-dipole interactions which can be controlled by an external electric field.  These interactions are repulsive when two molecules collide in a plane perpendicular to their oriented dipoles, but become attractive when the two molecules collide along their orientation axis [see e.g.~Eq.~\eqref{eq:DDI0}].  In addition, many of the experimentally relevant molecular species, including half of the alkali dimers~\cite{Zuchowski_Hutson}, are chemically reactive.  Hence, in order to avoid rapid losses from chemical reactions, these molecules must be confined in a reduced-dimensional geometry that energetically suppresses the reactive collisions by geometrically allowing only repulsive interactions.  In Ref.~\citen{deMiranda2011} it was shown that trapping molecules whose dipoles were aligned by an external electric field in a quasi-2D geometry generated by a 1D optical lattice reduced the chemical reaction rate by two orders of magnitude.  Chemical reactions between molecules can be further suppressed by loading the molecules into a 3D optical lattice.  In this case, the suppression  either comes from freezing out the motion of the molecules altogether or through the quantum Zeno mechanism~\cite{yan:observation_2013,PhysRevLett.112.070404}.  For the discussion of effective quantum magnetism in Sec.~\ref{sec:QMMol} we considered the case of molecules pinned in a deep optical lattice where only the internal degrees of freedom are relevant for the dynamics.  Many experiments, including the JILA KRb experiment, form molecules directly in a 3D lattice by assembling lattice-confined atoms, and this scenario is particularly relevant for them.

\subsection{Molecular structure and progress cooling complex molecules}
\label{sec:MolecularStructure}
Because the Coulomb interaction is isotropic, the electronic state of an atom can be expressed in terms of its conserved angular momenta.  One way to label atoms' electronic states is through term symbols $^{2S+1}L_J$, in which, e.~g., $\hat{\mathbf{J}}^2|JM\rangle=J\left(J+1\right)|JM\rangle$, with $|JM\rangle$ a state with $J$ quanta of angular momentum and a projection $M$ of $\mathbf{J}$ along a space-fixed axis.  Unlike atoms, which have spherical symmetry, (diatomic) molecules have only cylindrical symmetry with respect to the internuclear axis.  This means that the relevant quantities for classifying molecules are not the "length" of the angular momentum vectors, e.g.~$J$, but rather the projections of these vectors onto the internuclear axis.  For example, the projection of the total electronic angular momentum along this axis, $\Omega$, plays a similar role for molecules as the total angular momentum $J$ for atoms.  Table~\ref{table:terms} overviews the spectroscopic term notation used for molecules and compares it with the notation for atoms.  The appearance of an angular momentum is not meant to imply that this angular momentum is in fact a good quantum number.  In atoms, relativistic effects couple the electronic orbital angular momentum $\mathbf{L}$ to the electron spin $\mathbf{S}$, and so $L$ and $S$ are no longer good quantum numbers.  Similarly, the angular momenta of a molecule can be coupled in a variety of ways by the intramolecular interactions and so the projections, e.g.~$\Omega$, may not be good quantum numbers.  The various coupling schemes of the internal angular momenta of a molecule are classified by Hund's cases -- for details, see e.g.~Ref.~\citen{RSODM}.

\renewcommand{\arraystretch}{1.2}
\begin{table}[h]
\tbl{Labeling of electronic angular momenta for atoms and molecules}
{\begin{tabular}{|c|c|c|c|c|}
\hline 
&\multicolumn{2}{|c|}{Atomic species}&\multicolumn{2}{|c|}{Molecular species}\\
\hline 
Angular momentum&Notation&Values&Notation&Values\\
\hline
Orbital angular momentum&L&S,P,D,F,\dots&$\Lambda$&$\Sigma$, $\Pi$, $\Delta$,$\Phi$, \dots\\
\hline 
Spin angular momentum&S&0, ${1}/{2}$, $1$,\dots&$\Sigma$&0, ${1}/{2}$, $1$,\dots\\
\hline 
Total angular momentum&J& 0,${1}/{2}$, $1$, \dots&$\Omega$&0,$\pm {1}/{2}$, $\pm 1$, \dots\\
\hhline{|=|=|=|=|=|}
Term symbol& \multicolumn{2}{|c|}{$^{2S+1}L_J$}& \multicolumn{2}{|c|}{$^{2S+1}|\Lambda|_{|\Omega|}$}\\
\hline 
\end{tabular}}
\label{table:terms}
\end{table}

It is an experimental fact that the majority of diatomic molecules have $^1\Sigma$ ground states, which is to say that the electronic wave function is invariant with respect to all symmetry transformations (e.g.~rotations and reflections) of the molecule and the total spin is zero.  In particular, the alkali metal dimers, including the species KRb~\cite{Ni_Ospelkaus_08,Ospelkaus_Peer_08,Ospelkaus_Ni_09}, LiCs~\cite{PhysRevA.87.010701,PhysRevA.87.010702}, RbCs~\cite{Cho_McCarron_11,C1CP21769K,PhysRevA.85.032506,1405.6037}, NaK~\cite{PhysRevA.85.051602,PhysRevLett.109.085301}, LiNa~\cite{PhysRevA.86.021602}, and LiRb~\cite{PhysRevA.89.020702} which are under active experimental investigation, all have $^1\Sigma$ ground states.  Molecules that are formed from one open-shell atom, such as an alkali metal atom, and one closed-shell atom, such as an alkaline earth atom, have $^2\Sigma$ ground states due to the additional unpaired electron spin.  Examples of such molecules being pursued in experiment include RbSr~\cite{PhysRevA.88.023601}, RbYb~\cite{PhysRevA.88.052708}, and LiYb~\cite{PhysRevLett.106.205304, PhysRevA.87.013615}.  Additionally, many molecules which have highly diagonal Franck-Condon factors and so are amenable to direct laser cooling, such as SrF~\cite{Shuman2010}, YO~\cite{YO}, and the alkaline-earth monohydrides~\cite{DiRosa} have $^2\Sigma$ ground states\footnote{Laser cooling does not require a molecule with a $^2\Sigma$ ground state; $^1\Sigma$ ground states with highly diagonal Franck-Condon factors would be preferred in principle due to their simpler internal structure.  However, all $^1\Sigma$ species identified to have highly diagonal Franck-Condon factors have transition frequencies which are challenging to accommodate with currently available laser technology.~\cite{BarryThesis}}.  Finally, free radicals such as ClO, BrO, CH, NO, and OH all have $^2\Pi$ ground states which arise due to significant non-adiabatic effects in the electronic structure.  The hydroxyl radical OH, which was the first radical to be studied with microwave spectroscopy~\cite{PhysRev.100.1735}, has been shown to be amenable to evaporative cooling~\cite{OH}, and experiments are underway to cool it to quantum degeneracy.  While comparatively little work has been focused on cooling molecules with more than two atoms to degeneracy, a notable exception features polyatomic molecules which have both cylindrical symmetry and a small inversion splitting, known as symmetric top molecules.  Such molecules, for example methyl fluoride, CH$_3$F, display a very strong coupling to external electric fields which allows them to be cooled in a modified Sisyphus scheme known as opto-electrical cooling~\cite{PhysRevLett.107.263003,oec} combined with guided deceleration~\cite{PhysRevLett.112.013001}.

\section{Future prospects}
\label{sec:PartIII}

In this section we look forward to the future of quantum magnetic phenomena manifested in ultracold molecule experiments and identify both immediately available avenues as well as longer term prospects.  The vista can be visualized in the quantum simulation landscape shown in Fig.~\ref{fig:complexity}.  One immediate direction is to add an electric field to the Ramsey-type experiments described in Sec.~\ref{sec:Dynamical}.  An electric field enables the study of tunable XXZ magnetism rather than the spin-exchange (i.e.~XX) magnetism observed in present experiments via the introduction of nonzero $d_{\uparrow}$ and $d_{\downarrow}$ dipole moments, see Fig.~\ref{fig:Efield}.  In addition to tuning the $J_{\perp}$ and $J_z$ terms, which can drive a quantum phase transition from a (anti-) ferromagnetic phase to a paramagnetic phase, the electric field also introduces density-density interactions [the $V$ terms in Eq.~\eqref{eq:HijSpin}] and, more interestingly, density-spin couplings [$W$ terms in Eq.~\eqref{eq:HijSpin}] which do not appear for magnetism generated by superexchange, see Eq.~\eqref{sec:superexchange}.  The electric field can also aid the experiment in practical ways.  For example, at a certain "magic" value of the electric field, the polarizabilities of the $|0,0\rangle$ and $|1,0\rangle$ state coincide~\cite{Kotochigova_Demille_10} and the two states feel identical optical lattice potentials.  This assists in reducing decoherence similar to "magic" wavelength conditions in atomic clocks~\cite{Ye27062008}.  Finally, as an electric field mixes rotational levels of different parity, it introduces small ($\mathcal{O}\left(\beta_{\mathrm{DC}}\right)$ and higher) transition dipole matrix elements between states whose rotational quantum numbers differ by more than one.  An electric field hence allows for higher rotational states to be populated with direct, single-photon microwave transitions, a key step towards harnessing the full richness of the molecule's internal structure.

Another near-term experimental enhancement which enables exciting new physics is the introduction of local state preparation and probes.  As an example, one can imagine that only a specific, localized region of molecules are transferred to the rotational excited state.  This localized excitation will propagate throughout the system via the dipolar spin-exchange process described by Eq.~\eqref{eq:dipolarXXZ}, and high-resolution imaging could be used to track this spin transport as well as the spread of correlations.  The propagation of correlations in a long-range interacting system can display fundamentally different behavior from short-range interacting systems, such as the breakdown of locality~\cite{1367-2630-15-8-083007,PhysRevX.3.031015,PhysRevLett.111.207202,PhysRevLett.111.260401,1401.5088,1401.5387}.  In addition, such long-range spin transport has strong analogies to the transport of energy in light-harvesting complexes and exciton transport, leading to diverse quantum simulation prospects~\cite{Guenter}.  In the study of spin transport phenomena, the disordered filling of the lattice due to the relatively hot temperature of the molecules can be advantageous, as disordered, long-range interacting spin systems have been identified as prime candidates for observing many-body localization (MBL)~\cite{basko:many-body_2006,pal:many-body_2010,yao:many-body_2013,kwasigroch:bose-einstein_2013,serbyn:interferometric_2014}.  A system displaying MBL transports neither heat nor charge, even when the amount of energy injected is macroscopic.  MBL is currently of great interest, in part because a system displaying MBL can break continuous symmetries or display topological order in situations where such order is usually forbidden~\cite{pal:many-body_2010,Chandran_etal}.  Hence, introducing local probes to present ultracold molecule experiments immediately provides access to a wealth of new phenomena.

A third direction for future experiments is to decrease the temperature of the molecules loaded into an optical lattice, which will in turn increase the lattice filling and allow for the study of quantum magnetism in equilibrium.  Here, we mention only a few examples of novel physics, and refer the interested reader to recent review articles which focus on the equilibrium many-body physics of polar molecules~\cite{doi:10.1021/cr2003568}.  A interesting feature of long-range interacting spin models is that they can display features normally associated with gapless, quantum critical phases, such as algebraic decay of correlations and large (area-law violating~\cite{RevModPhys.82.277}) entanglement entropy, even in gapped phases~\cite{PRA_72_063407,PRL_97_150404,Hauke,PRL_109_267203,LRKitaev}.  Additionally, the interplay between coherent tunneling, spin interactions and chemical reactions~\cite{yan:observation_2013,PhysRevLett.112.070404} will allow us  to study itinerant quantum magnetism such as generalized $t-J$ models~\cite{Gorshkov_Manmana_11,Gorshkov_Manmana_11b}.  This latter point is relevant to many of the earlier examples given in this section, as little is known about the behavior of, e.g., a many-body localization transition in the presence of itineracy.

\begin{figure}[bp]
\centerline{\includegraphics[width=0.666\columnwidth]{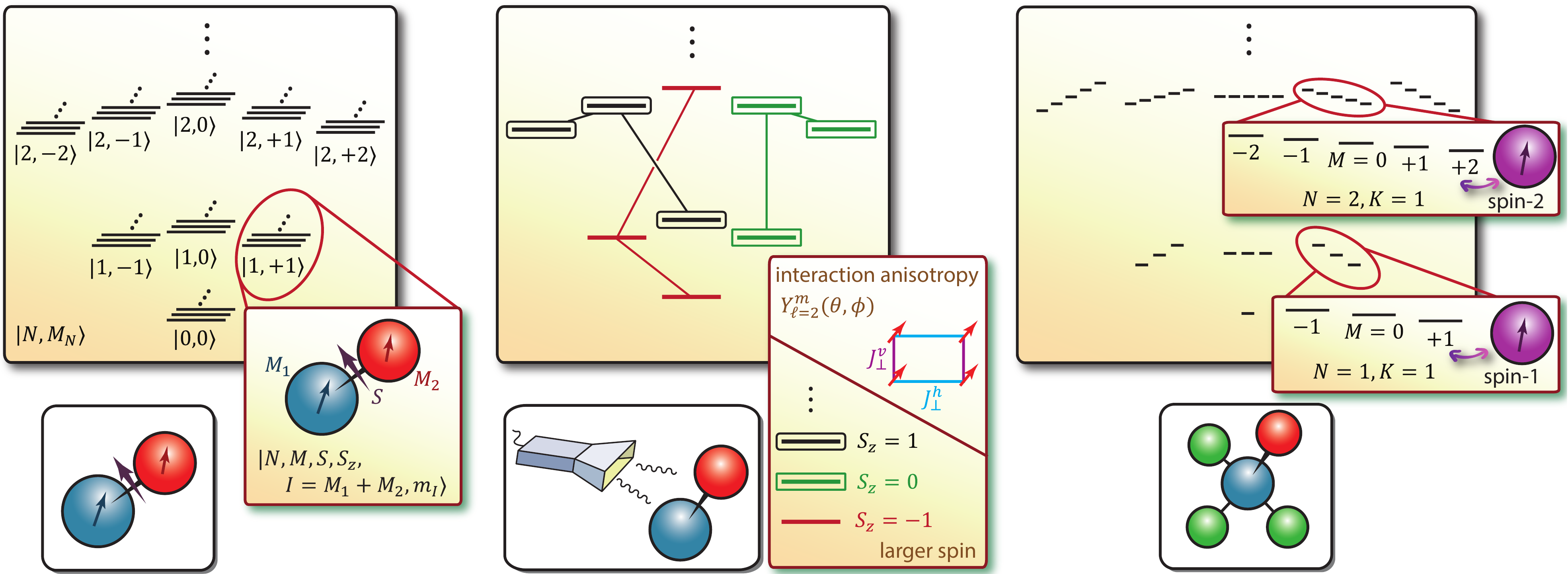}}
\caption{\label{fig:complexermolecules}  More complex physics from more complex molecules: beyond the rigid rotor. Left: most molecules have many hyperfine states, the product of the  hyperfine degrees of freedom of the constituent atoms. ($^{40}$K$^{87}$Rb, for example, has 36 hyperfine states.)  These degrees of freedom may serve to realize, for example, multi-orbital spin models.  Middle: microwaves used to generate dressed states as the effective spin degrees of freedom allow the manipulation of the dipolar interaction in ways that do not occur for bare rotational states.  This allows much greater control of both the spatial and spin anisotropy in the interaction.  Right: molecules with more than two atoms give rise to new rotational states due to their ability to rotate around more than one axis. (The energy splittings are not to scale, and the magnitude of the "small" grouped splittings may even be larger than the large separations in some molecules.) These allow polar molecules to effectively behave as magnetic dipoles. Some possible physics arising from these more complex situations is depicted in Fig.~\ref{fig:complexity}. }
\end{figure}

A final, longer-term direction for ultracold molecule realizations of quantum magnetism is to increase molecular diversity, both in the form of many different species with the same essential molecular structure as well as molecules with more complex structure.  As an example of the former point, all of the alkali metal dimer molecules have $^1\Sigma$ ground states and hence essentially the same molecular structure, but differ in terms of energy scales: the dipole moments of the alkali dimers vary over an order of magnitude, as well as the degree over which they can be polarized by an electric field of a given strength~\cite{PhysRevA.88.023605}.  In addition, half of the alkali dimers are chemically reactive, while the other half are not~\cite{Zuchowski_Hutson}.  An increase in the complexity of the internal structure beyond two rotational levels of a rigid rotor can be realized either by considering molecules with more complex structure or by accessing additional levels with more complex microwave/optical dressing schemes; we display both possibilities along the complexity axis of Fig.~\ref{fig:complexity}.  These possibilities are expounded upon further in Fig.~\ref{fig:complexermolecules}.  In the far left panel, we display that even $\Sigma$-state molecules whose rotational structure is that of a simple rigid rotor often have other internal degrees of freedom with energies much smaller than rotational scales.  In $^1\Sigma$ molecules such as the alkali dimers, the internal structure consists of hyperfine degrees of freedom which can be quite numerous; $^{40}$K$^{87}$Rb has a hyperfine degeneracy of 36.  The hyperfine structure is coupled to the rotational structure for nuclear spins $I\ge 1$ by nuclear quadrupole couplings.  Such couplings allow for controllable population of nuclear spin states through microwave transitions~\cite{Ospelkaus_Ni_10}.  Additional internal structure in $\Sigma$ state molecules can also arise from unpaired electron spin, such as $^2\Sigma$ states for alkali metal-alkaline earth diatomics, long-lived $^3\Sigma$ excited states of many $^1\Sigma$ ground state molecules, and even exotic high-spin states such as the $^6\Sigma$ state of CrRb.  Such additional degrees of freedom may be useful in simulating multi-orbital models with ultracold molecules, and can also be used to design complex spin models~\cite{Brennen2,Brennen2007}.

Dressing of molecules with microwave or optical fields~\cite{PhysRevLett.103.155302,Gorshkov_Manmana_11,Gorshkov_Manmana_11b} leads to a very rich landscape of possibilities by accessing the full tensorial structure and anisotropy of the dipole-dipole interaction, Eq.~\eqref{eq:DDI}, opening up components beyond the $q=0$ one in Eq.~\eqref{eq:DDI0}.  This enables the study of compass-type models in which the coupling strength depends on the spatial direction of the coupling, which has been proposed to lead to symmetry protected topological phases~\cite{PhysRevB.87.081106} and models with true topological order such as the Kitaev honeycomb model~\cite{doi:10.1080/00268976.2013.800604}.  In addition, the full tensorial structure of the dipole-dipole interaction can exchange internal rotational angular momentum with orbital angular momentum, and hence lead to spin-orbit coupled rotational excitations which are chiral and feature a non-trivial Berry phase~\cite{Chiron}.  The essential idea of the microwave dressing procedure is to both (1) isolate $d$-level systems comprised of rotational levels linked by near-resonant radiation and (2) tune the interactions between these levels by choosing the polarizations, field strengths, and frequencies of the dressing radiation.  Microwave dressing enables more complex interactions because many rotational levels can be coherently superposed, see Fig.~\ref{fig:complexermolecules}, meaning that many dipole matrix elements are involved.  Additionally, provided that such superpositions can be efficiently prepared, many such "dressed states" can be considered, leading to effectively high-dimensional spin representations, e.g. the~spin-1 realized by the black, green, and red dressed levels in Fig.~\ref{fig:complexermolecules}.  Dressing with optical radiation which has spatial variation on the lattice scale enables even more  diverse phenomena, such as topological flat bands~\cite{PhysRevLett.109.266804} and fractional Chern insulators~\cite{PhysRevLett.110.185302}.

Finally, new quantum magnetic phenomena arise in molecules with more complex internal structure than diatomic molecules with $\Sigma$ ground states.  As one example, we mention \emph{symmetric top molecules} (STMs), molecules with (1) cylindrical symmetry and (2) nonzero rotational or orbital angular momentum projection along the body frame of the molecule.  While diatomic $\Sigma$ state molecules are cylindrically symmetric, they rotate about an axis perpendicular to their body frame, and hence do not satisfy (2).  An example of an STM is the polyatomic molecule methyl fluoride, {\MF}, shown in Fig.~\ref{fig:complexermolecules}.  In methyl fluoride the rotational angular momentum can have some projection $K$ onto the body frame of the molecule, and this leads to the rotational spectrum shown in the right panel of Fig.~\ref{fig:complexermolecules}.  In contrast to the Stark spectrum of $\Sigma$-state molecules shown in the left and center panels of Fig.~\ref{fig:complexermolecules}, STMs display a linear coupling to an external electric field.  This allows STMs to behave like elemental magnetic dipoles, leading to a large number of quantum simulation prospects~\cite{wall2013simulating}.  For example, STMs in static fields can realize models of arbitrarily large integer spins interacting through anisotropic, long-range interactions~\cite{wall2013simulating}.  We note that STMs do not have to contain more than two atoms; diatomic molecules with orbital angular momentum projection $|\Lambda|>0$ such as the $^2\Pi$ ground state of OH~\cite{stuhl2012evaporative} also behave as STMs in modest electric fields.

\section{Conclusions}
\label{sec:Concl}

In conclusion, we have shown how to realize effective quantum magnetism with polar molecules pinned in optical lattices.  The basic idea is to  encode an effective spin 1/2 degree of freedom in two isolated rotational levels and  to generate spin interactions via dipolar interactions.  The effective spin-spin couplings even in the simplest scenario can be tuned by an external electric field as well as by the choice of rotational states forming the effective spin-1/2.  This way of realizing quantum magnetism was contrasted with the superexchange mechanism.  In the latter the spin-spin couplings are induced via short-range interactions through virtual tunneling processes and hence  the  motional temperature plays an essential role.  In polar molecule quantum magnetism, motional and spin temperatures are decoupled, and so quantum magnetism can be seen even in hot gases as long as a well characterized initial spin state is   prepared.  We reviewed the experimental confirmation of dipolar quantum magnetism in optical lattice experiments with KRb molecules at JILA, and discussed the new theoretical tools that were developed to verify and explain the experimental results.  We expect that these tools, as well as the general theme that far from equilibrium dynamics can produce correlated quantum many-body physics in long-range interacting gases without motional degeneracy, will be of use in the cold gases community.  Finally, we suggest some directions for future realizations of quantum magnetism with ultracold molecules, ranging from near-term experimental upgrades to long-term realizations of molecules with greater complexity.

\section*{Acknowledgments}

We would like to acknowledge conversations with Lincoln Carr, Michael Foss-Feig, Alexey V.~Gorshkov, Kenji Maeda, Salvatore Manmana, Alex Pikovski, Johannes Schachenmayer, Bihui Zhu, and especially the countless discussions with the JILA KRb experimental group over the last few years: Debbie Jin, Jun Ye, Bryce Gadway, Bo Yan, Jacob Covey, Brian Neyenhuis, and Steven Moses. This work was supported by the NSF (PIF-1211914 and PFC-1125844), AFOSR and ARO individual investigator awards, and the ARO with funding from the DARPA-OLE program. The authors also  thank the the Aspen Center for Physics and KITP (NSF-PHY11-25915). KRAH and MLW thank the NRC for support.


\begin{thebibliography}{100}

\bibitem{anderson1995}
M.~H. Anderson, J.~R. Ensher, M.~R. Matthews, C.~E. Wieman, and E.~A. Cornell,
  Science {\bf 269},  198  (1995).

\bibitem{davis1995}
K.~B. Davis, M.-O. Mewes, M.~R. Andrews, N.~J. {van Druten}, D.~S. Durfee,
  D.~M. Kurn, and W. Ketterle, Phys. Rev. Lett. {\bf 75},  3969  (1995).

\bibitem{bradleyCC1995}
C.~C. Bradley, C.~A. Sackett, J.~J. Tollett, and R.~G. Hulet, Phys. Rev. Lett.
  {\bf 75},  1687  (1995).

\bibitem{PhysRevLett.83.2498}
M.~R. Matthews, B.~P. Anderson, P.~C. Haljan, D.~S. Hall, C.~E. Wieman, and
  E.~A. Cornell, Phys. Rev. Lett. {\bf 83},  2498  (1999).

\bibitem{PhysRevLett.86.2926}
B.~P. Anderson, P.~C. Haljan, C.~A. Regal, D.~L. Feder, L.~A. Collins, C.~W.
  Clark, and E.~A. Cornell, Phys. Rev. Lett. {\bf 86},  2926  (2001).

\bibitem{Bloch}
I. Bloch, J. Dalibard, and W. Zwerger, Rev. Mod. Phys. {\bf 80},  885  (2008).

\bibitem{DeMarco}
B. DeMarco and D.~S. Jin, Science {\bf 285},  1703  (1999).

\bibitem{AEAs}
M.~A. Cazalilla and A.~M. Rey, arXiv:1403.2792  .

\bibitem{PhysRevLett.108.215301}
M. Lu, N.~Q. Burdick, and B.~L. Lev, Phys. Rev. Lett. {\bf 108},  215301
  (2012).

\bibitem{PhysRevLett.107.190401}
M. Lu, N.~Q. Burdick, S.~H. Youn, and B.~L. Lev, Phys. Rev. Lett. {\bf 107},
  190401  (2011).

\bibitem{PhysRevLett.108.210401}
K. Aikawa, A. Frisch, M. Mark, S. Baier, A. Rietzler, R. Grimm, and F.
  Ferlaino, Phys. Rev. Lett. {\bf 108},  210401  (2012).

\bibitem{bloch2005ultracold}
I. Bloch, Nature Physics {\bf 1},  23  (2005).

\bibitem{feynmanRP1982}
R.~P. Feynman, Int. J. Theor. Phys. {\bf 21},  467  (1982).

\bibitem{Greiner_Mandel_02}
M. Greiner, O. Mandel, T. Esslinger, T.~W. Hansch, and I. Bloch, Nature {\bf
  415},  39  (2002).

\bibitem{jordens2008mott}
R. J{\"o}rdens, N. Strohmaier, K. G{\"u}nter, H. Moritz, and T. Esslinger,
  Nature {\bf 455},  204  (2008).

\bibitem{schneider:metallic_2008}
U. Schneider, L. Hackermüller, S. Will, T. Best, I. Bloch, T.~A. Costi, R.~W.
  Helmes, D. Rasch, and A. Rosch, Science {\bf 322},  1520  (2008).

\bibitem{Auerbach}
A. Auerbach, {\em Interacting {E}lectrons and {Q}uantum {M}agnetism} (Springer,
  Berlin, 1994).

\bibitem{sachdev_quantum_2008}
S. Sachdev, Nature Physics {\bf 4},  173  (2008).

\bibitem{lacroix_introduction_2011}
{\em Introduction to frustrated magnetism}, edited by C. Lacroix, P. Mendels,
  and F. Mila (Springer, Heidelberg, 2011).

\bibitem{barnett:quantum_2006}
R. Barnett, D. Petrov, M. Lukin, and E. Demler, Phys. Rev. Lett. {\bf 96},
  190401  (2006).

\bibitem{Wall_PRA_2010}
M.~L. Wall and L.~D. Carr, Phys. Rev. A {\bf 82},  013611  (2010).

\bibitem{Gorshkov_Manmana_11}
A.~V. Gorshkov, S.~R. Manmana, G. Chen, J. Ye, E. Demler, M.~D. Lukin, and
  A.~M. Rey, Phys. Rev. Lett. {\bf 107},  115301  (2011).

\bibitem{Gorshkov_Manmana_11b}
A.~V. Gorshkov, S.~R. Manmana, G. Chen, E. Demler, M.~D. Lukin, and A.~M. Rey,
  Phys. Rev. A {\bf 84},  033619  (2011).

\bibitem{Zare}
R. Zare, {\em Angular Momentum: Understanding Spatial Aspects in Chemistry and
  Physics} (Wiley, New York, 1988).

\bibitem{Jaksch_PRL_1998}
D. Jaksch, C. Bruder, J.~I. Cirac, C.~W. Gardiner, and P. Zoller, Phys. Rev.
  Lett. {\bf 81},  3108  (1998).

\bibitem{Duan}
L.-M. Duan, E. Demler, and M.~D. Lukin, Phys. Rev. Lett. {\bf 91},  090402
  (2003).

\bibitem{PhysRevA.84.061605}
A. Pikovski, M. Klawunn, A. Recati, and L. Santos, Phys. Rev. A {\bf 84},
  061605  (2011).

\bibitem{hazzard:quantum-correlations_2014}
K.~R.~A. Hazzard, M. van~den Worm, M. Foss-Feig, S.~R. Manmana, E.~D. Torre, T.
  Pfau, M. Kastner, and A.~M. Rey, arXiv:1406.0937  .

\bibitem{PhysRevA.87.042101}
M. Foss-Feig, K.~R.~A. Hazzard, J.~J. Bollinger, and A.~M. Rey, Phys. Rev. A
  {\bf 87},  042101  (2013).

\bibitem{1367-2630-15-8-083007}
M. van~den Worm, B.~C. Sawyer, J.~J. Bollinger, and M. Kastner, New J. Phys.
  {\bf 15},  083007  (2013).

\bibitem{1367-2630-15-11-113008}
M. Foss-Feig, K.~R.~A. Hazzard, J.~J. Bollinger, A.~M. Rey, and C.~W. Clark,
  New J. Phys. {\bf 15},  113008  (2013).

\bibitem{hazzard:far-from-equilibrium_2013}
K.~R.~A. Hazzard, S.~R. Manmana, M. Foss-Feig, and A.~M. Rey, Phys. Rev. Lett.
  {\bf 110},  075301  (2013).

\bibitem{yan:observation_2013}
B. Yan, S.~A. Moses, B. Gadway, J.~P. Covey, K.~R.~A. Hazzard, A.~M. Rey, D.~S.
  Jin, and J. Ye, Nature {\bf 501},  521  (2013).

\bibitem{neyenhuis:anisotropic_2012}
B. Neyenhuis, B. Yan, S.~A. Moses, J.~P. Covey, A. Chotia, A. Petrov, S.
  Kotochigova, J. Ye, and D.~S. Jin, Phys. Rev. Lett. {\bf 109},  230403
  (2012).

\bibitem{hazzard:many-body_2014}
K.~R.~A. Hazzard, B. Gadway, M. Foss-Feig, B. Yan, S.~A. Moses, J.~P. Covey,
  N.~Y. Yao, M.~D. Lukin, J. Ye, D.~S. Jin, and A.~M. Rey, arXiv:1402.2354  .

\bibitem{waugh:approach_1968}
J.~S. Waugh, L.~M. Huber, and U. Haeberlen, Phys. Rev. Lett. {\bf 20},  180
  (1968).

\bibitem{johnson:observation_1984}
B.~R. Johnson, J.~S. Denker, N. Bigelow, L.~P. L\'evy, J.~H. Freed, and D.~M.
  Lee, Phys. Rev. Lett. {\bf 52},  1508  (1984).

\bibitem{lemke:p-wave_2011}
N.~D. Lemke, J. von Stecher, J.~A. Sherman, A.~M. Rey, C.~W. Oates, and A.~D.
  Ludlow, Phys. Rev. Lett. {\bf 107},  103902  (2011).

\bibitem{martin:quantum_2013}
M.~J. Martin, M. Bishof, M.~D. Swallows, X. Zhang, C. Benko, J. von Stecher,
  A.~V. Gorshkov, A.~M. Rey, and J. Ye, Science {\bf 341},  632  (2013).

\bibitem{stoneman:resonant-collision_1987}
R.~C. Stoneman, M.~D. Adams, and T.~F. Gallagher, Phys. Rev. Lett. {\bf 58},
  1324  (1987).

\bibitem{demarco:spin_2002}
B. DeMarco and D.~S. Jin, Phys. Rev. Lett. {\bf 88},  040405  (2002).

\bibitem{koschorreck:universal_2013}
M. Koschorreck, D. Pertot, E. Vogt, and M. Kohl, Nature Physics {\bf 9},  405
  (2013).

\bibitem{bardon:transverse_2013}
A.~B. Bardon, S. Beattie, C. Luciuk, W. Cairncross, D. Fine, N.~S. Cheng,
  G.~J.~A. Edge, E. Taylor, S. Zhang, S. Trotzky, and J.~H. Thywissen, Science
  {\bf 344},  722  (2014).

\bibitem{krauser:giant_2014}
J.~S. Krauser, U. Ebling, N. Fl\"{a}schner, J. Heinze, K. Sengstock, M.
  Lewenstein, A. Eckardt, and C. Becker, Science {\bf 343},  157  (2014).

\bibitem{white:density_1992}
S.~R. White, Phys. Rev. Lett. {\bf 69},  2863  (1992).

\bibitem{vidal:efficient_2004}
G. Vidal, Phys. Rev. Lett. {\bf 93},  040502  (2004).

\bibitem{daley:time-dependent_2004}
A.~J. Daley, C. Kollath, U. Schollw{\"o}ck, and G. Vidal, J. Stat. Mech.
  P04005  (2004).

\bibitem{schollwoeck:density-matrix_2005}
U. Schollw\"ock, Rev. Mod. Phys. {\bf 77},  259  (2005).

\bibitem{schollwoeck:density-matrix_2011}
U. Schollw{\"o}ck, Annals of Physics {\bf 326},  96   (2011).

\bibitem{doi:10.1146/annurev-conmatphys-020911-125018}
E. Stoudenmire and S.~R. White, Annual Review of Condensed Matter Physics {\bf
  3},  111  (2012).

\bibitem{abrikosov_methods_1975}
A.~A. Abrikosov, {\em Methods of Quantum Field Theory in Statistical Physics},
  revised ed. (Dover Publications, New York, USA, 1975).

\bibitem{negele_quantum_1998}
J.~W. Negele and H. Orland, {\em Quantum {Many-Particle} Systems} (Westview
  Press, USA, 1998).

\bibitem{polkovnikov:phase_2010}
A. Polkovnikov, Annals of Physics {\bf 325},  1790   (2010).

\bibitem{noack:diagonalization_2005}
R.~M. Noack and S.~R. Manmana, AIP Conference Proceedings {\bf 789},  93
  (2005).

\bibitem{basko:many-body_2006}
D.~M. Basko, I.~L. Aleiner, and B.~L. Altshuler, Annals of Physics {\bf 321},
  1126  (2006).

\bibitem{pal:many-body_2010}
A. Pal and D.~A. Huse, Phys. Rev. B {\bf 82},  174411  (2010).

\bibitem{mathey:dynamics_2011}
L. Mathey, K.~J. G{\"u}nter, J. Dalibard, and A. Polkovnikov, arXiv:1112.1204
  .

\bibitem{mathey:light_2010}
L. Mathey and A. Polkovnikov, Phys. Rev. A {\bf 81},  033605  (2010).

\bibitem{yao:many-body_2013}
N.~Y. Yao, C.~R. Laumann, S. Gopalakrishnan, M. Knap, M. Mueller, E.~A. Demler,
  and M.~D. Lukin, arXiv:1311.7151  .

\bibitem{kwasigroch:bose-einstein_2013}
M.~P. Kwasigroch and N.~R. Cooper, arXiv:1311.5393  .

\bibitem{serbyn:interferometric_2014}
M. Serbyn, M. Knap, S. Gopalakrishnan, Z. Papi{\'c}, N.~Y. Yao, C.~R. Laumann,
  D.~A. Abanin, M.~D. Lukin, and E.~A. Demler, arXiv:1403.0693  .

\bibitem{foelling:quantum_2014}
S. F{\"o}lling, arXiv:1403.6842  .

\bibitem{rey:probing_2014}
A. Rey, A. Gorshkov, C. Kraus, M. Martin, M. Bishof, M. Swallows, X. Zhang, C.
  Benko, J. Ye, N. Lemke, and A. Ludlow, Annals of Physics {\bf 340},  311
  (2014).

\bibitem{knap:probing_2013}
M. Knap, A. Kantian, T. Giamarchi, I. Bloch, M.~D. Lukin, and E. Demler, Phys.
  Rev. Lett. {\bf 111},  147205  (2013).

\bibitem{Bistritzer:intrinsic_2007}
R. Bistritzer and E. Altman, Proc. Natl. Acad. Sci. {\bf 104},  9955  (2007).

\bibitem{degrandi:quench_2010}
C. {De Grandi}, V. Gritsev, and A. Polkovnikov, Phys. Rev. B {\bf 81},  012303
  (2010).

\bibitem{degrandi:quench-sine-gordon_2010}
C. {De Grandi}, V. Gritsev, and A. Polkovnikov, Phys. Rev. B {\bf 81},  224301
  (2010).

\bibitem{gritsev:universal_2010}
V. Gritsev and A. Polkovnikov,  in {\em Understanding in Quantum Phase
  Transitions}, edited by L. Carr (Taylor \& Francis, 2010, 2010).

\bibitem{mitra:time_2012}
A. Mitra, Phys. Rev. Lett. {\bf 109},  260601  (2012).

\bibitem{mitra:thermalization_2012}
A. Mitra and T. Giamarchi, Phys. Rev. B {\bf 85},  075117  (2012).

\bibitem{mitra:mode-coupling_2011}
A. Mitra and T. Giamarchi, Phys. Rev. Lett. {\bf 107},  150602  (2011).

\bibitem{polkovnikov:nonequilibrium_2011}
A. Polkovnikov, K. Sengupta, A. Silva, and M. Vengalattore, Rev. Mod. Phys.
  {\bf 83},  863  (2011).

\bibitem{karrasch:luttinger-liquid_2012}
C. Karrasch, J. Rentrop, D. Schuricht, and V. Meden, Phys. Rev. Lett. {\bf
  109},  126406  (2012).

\bibitem{vosk:many-body_2012}
R. Vosk and E. Altman, arXiv:1205.0026  .

\bibitem{dallatorre:dynamics_2012}
E.~G. {Dalla Torre}, E. Demler, T. Giamarchi, and E. Altman, Phys. Rev. B {\bf
  85},  184302  (2012).

\bibitem{dallatorre:universal_2013}
E.~G. {Dalla Torre}, E. Demler, and A. Polkovnikov, Phys. Rev. Lett. {\bf 110},
   090404  (2013).

\bibitem{sarkar:perturbative_2013}
S. {De Sarkar}, R. Sensarma, and K. Sengupta, arXiv:1308.4689  .

\bibitem{mitra:correlation_2013}
A. Mitra, Phys. Rev. B {\bf 87},  205109  (2013).

\bibitem{acevedo:new_2014}
O. Acevedo, L. Quiroga, J. Rodr\'iguez, F.\, and F. Johnson, N.\, Phys. Rev.
  Lett. {\bf 112},  030403  (2014).

\bibitem{Quemener_Julienne_12}
G. Qu\'{e}m\'{e}ner and P.~S. Julienne, Chemical Reviews {\bf 112},  4949
  (2012).

\bibitem{Stwalley}
J.~T. Bahns, W.~C. Stwalley, and P.~L. Gould, The Journal of Chemical Physics
  {\bf 104},  9689  (1996).

\bibitem{DiRosa}
M. Rosa, The European Physical Journal D - Atomic, Molecular, Optical and
  Plasma Physics {\bf 31},  395  (2004).

\bibitem{Ni_Ospelkaus_08}
K.-K. Ni, S. Ospelkaus, M.~H.~G. {de Miranda}, A. Pe'{e}r, B. Neyenhuis, J.~J.
  Zirbel, S. Kotochigova, P.~S. Julienne, D.~S. Jin, and J. Ye, Science {\bf
  322},  231  (2008).

\bibitem{Shapiro}
E.~A. Shapiro, A. Pe'er, J. Ye, and M. Shapiro, Phys. Rev. Lett. {\bf 101},
  023601  (2008).

\bibitem{Ospelkaus_Ni_10}
S. Ospelkaus, K.-K. Ni, G. Qu\'em\'ener, B. Neyenhuis, D. Wang, M.~H.~G.
  de~Miranda, J.~L. Bohn, J. Ye, and D.~S. Jin, Phys. Rev. Lett. {\bf 104},
  030402  (2010).

\bibitem{Meerakker}
S.~Y.~T. van~de Meerakker, H.~L. Bethlem, N. Vanhaecke, and G. Meijer, Chemical
  Reviews {\bf 112},  4828  (2012).

\bibitem{Hutzler}
N.~R. Hutzler, H.-I. Lu, and J.~M. Doyle, Chemical Reviews {\bf 112},  4803
  (2012).

\bibitem{Carr_Demille_09}
L.~D. Carr, D. Demille, R.~V. Krems, and J. Ye, New J. Phys. {\bf 11},  055049
  (2009).

\bibitem{KochandShapiro}
C.~P. Koch and M. Shapiro, Chemical Reviews {\bf 112},  4928  (2012).

\bibitem{Koehler}
T. K\"ohler, K. G\'oral, and P.~S. Julienne, Rev. Mod. Phys. {\bf 78},  1311
  (2006).

\bibitem{Zuchowski_Hutson}
P.~S. \.{Z}uchowski and J.~M. Hutson, Phys. Rev. A {\bf 81},  060703  (2010).

\bibitem{deMiranda2011}
M.~H.~G. de~Miranda, A. Chotia, B. Neyenhuis, D. Wang, G. Qu\'em\'ener, S.
  Ospelkaus, J.~L. Bohn, J. Ye, and D.~S. Jin, Nature Physics {\bf 7},  502
  (2011).

\bibitem{PhysRevLett.112.070404}
B. Zhu, B. Gadway, M. Foss-Feig, J. Schachenmayer, M.~L. Wall, R.~A. Hazzard,
  K.\, B. Yan, A. Moses, S.\, P. Covey, J.\, S. Jin, D.\, J. Ye, M. Holland,
  and M. Rey, A.\, Phys. Rev. Lett. {\bf 112},  070404  (2014).

\bibitem{RSODM}
J. Brown and A. Carrington, {\em Rotational Spectroscopy of Diatomic Molecules}
  (Cambridge University Press, Cambridge, 2003).

\bibitem{Ospelkaus_Peer_08}
S. Ospelkaus, A. Pe'er, K.~K. Ni, J.~J. Zirbel, B. Neyenhuis, S. Kotochigova,
  P.~S. Julienne, J. Ye, and D.~S. Jin, Ultracold dense gas of deeply bound
  heteronuclear molecules, 2008.

\bibitem{Ospelkaus_Ni_09}
S. Ospelkaus, K.~K. Ni, M.~H.~G. {de Miranda}, A. Pe'er, B. Neyenhuis, D. Wang,
  S. Kotochigova, P.~S. Julienne, D.~S. Jin, and J. Ye, Faraday Discuss. {\bf
  142},  351  (2009).

\bibitem{PhysRevA.87.010701}
M. Repp, R. Pires, J. Ulmanis, R. Heck, E.~D. Kuhnle, M. Weidem\"uller, and E.
  Tiemann, Phys. Rev. A {\bf 87},  010701  (2013).

\bibitem{PhysRevA.87.010702}
S.-K. Tung, C. Parker, J. Johansen, C. Chin, Y. Wang, and P.~S. Julienne, Phys.
  Rev. A {\bf 87},  010702  (2013).

\bibitem{Cho_McCarron_11}
H. Cho, D. McCarron, D.~L. Jenkin, M.~P. K\"{o}ppinger, and S.~L. Cornish,
  European Physical Journal D  1  (2011), 10.1140/epjd/e2011-10716-1.

\bibitem{C1CP21769K}
M. Debatin, T. Takekoshi, R. Rameshan, L. Reichs\"{o}llner, F. Ferlaino, R.
  Grimm, R. Vexiau, N. Bouloufa, O. Dulieu, and H.-C. Nagerl, Phys. Chem. Chem.
  Phys. {\bf 13},  18926  (2011).

\bibitem{PhysRevA.85.032506}
T. Takekoshi, M. Debatin, R. Rameshan, F. Ferlaino, R. Grimm, H.-C. N\"agerl,
  C.~R. Le~Sueur, J.~M. Hutson, P.~S. Julienne, S. Kotochigova, and E. Tiemann,
  Phys. Rev. A {\bf 85},  032506  (2012).

\bibitem{1405.6037}
T. Takekoshi, L. Reichs\"{o}llner, A. Schindewolf, J.~M. Hutson, C.~R.~L.
  Sueur, O. Dulieu, F. Ferlaino, R. Grimm, and H.-C. N\"{a}gerl,
  arXiv:1405.6037  .

\bibitem{PhysRevA.85.051602}
J.~W. Park, C.-H. Wu, I. Santiago, T.~G. Tiecke, S. Will, P. Ahmadi, and M.~W.
  Zwierlein, Phys. Rev. A {\bf 85},  051602  (2012).

\bibitem{PhysRevLett.109.085301}
C.-H. Wu, J.~W. Park, P. Ahmadi, S. Will, and M.~W. Zwierlein, Phys. Rev. Lett.
  {\bf 109},  085301  (2012).

\bibitem{PhysRevA.86.021602}
M.-S. Heo, T.~T. Wang, C.~A. Christensen, T.~M. Rvachov, D.~A. Cotta, J.-H.
  Choi, Y.-R. Lee, and W. Ketterle, Phys. Rev. A {\bf 86},  021602  (2012).

\bibitem{PhysRevA.89.020702}
S. Dutta, J. Lorenz, A. Altaf, D.~S. Elliott, and Y.~P. Chen, Phys. Rev. A {\bf
  89},  020702  (2014).

\bibitem{PhysRevA.88.023601}
B. Pasquiou, A. Bayerle, S.~M. Tzanova, S. Stellmer, J. Szczepkowski, M.
  Parigger, R. Grimm, and F. Schreck, Phys. Rev. A {\bf 88},  023601  (2013).

\bibitem{PhysRevA.88.052708}
M. Borkowski, P.~S. $\dot{Z}$uchowski, R. Ciury\l{}o, P.~S. Julienne, D.
  K\k{e}dziera, L. Mentel, P. Tecmer, F. M\"unchow, C. Bruni, and A. G\"orlitz,
  Phys. Rev. A {\bf 88},  052708  (2013).

\bibitem{PhysRevLett.106.205304}
H. Hara, Y. Takasu, Y. Yamaoka, J.~M. Doyle, and Y. Takahashi, Phys. Rev. Lett.
  {\bf 106},  205304  (2011).

\bibitem{PhysRevA.87.013615}
A.~H. Hansen, A.~Y. Khramov, W.~H. Dowd, A.~O. Jamison, B. Plotkin-Swing, R.~J.
  Roy, and S. Gupta, Phys. Rev. A {\bf 87},  013615  (2013).

\bibitem{Shuman2010}
E.~S. Shuman, J.~F. Barry, and D. DeMille, Nature {\bf 467},  820  (2010).

\bibitem{YO}
M.~T. Hummon, M. Yeo, B.~K. Stuhl, A.~L. Collopy, Y. Xia, and J. Ye, Phys. Rev.
  Lett. {\bf 110},  143001  (2013).

\bibitem{BarryThesis}
J. Barry, Ph.D. thesis, Yale University, 2013.

\bibitem{PhysRev.100.1735}
G.~C. Dousmanis, T.~M. Sanders, and C.~H. Townes, Phys. Rev. {\bf 100},  1735
  (1955).

\bibitem{OH}
B.~K. Stuhl, M.~T. Hummon, M. Yeo, G. Qu\'em\'ener, J.~L. Bohn, and J. Ye,
  Nature {\bf 492},  396  (2012).

\bibitem{PhysRevLett.107.263003}
B.~G.~U. Englert, M. Mielenz, C. Sommer, J. Bayerl, M. Motsch, P.~W.~H. Pinkse,
  G. Rempe, and M. Zeppenfeld, Phys. Rev. Lett. {\bf 107},  263003  (2011).

\bibitem{oec}
M. Zeppenfeld, B.~G.~U. Englert, R. Glockner, A. Prehn, M. Mielenz, C. Sommer,
  L.~D. van Buuren, M. Motsch, and G. Rempe, Nature {\bf 491},  570  (2012).

\bibitem{PhysRevLett.112.013001}
S. Chervenkov, X. Wu, J. Bayerl, A. Rohlfes, T. Gantner, M. Zeppenfeld, and G.
  Rempe, Phys. Rev. Lett. {\bf 112},  013001  (2014).

\bibitem{Kotochigova_Demille_10}
S. Kotochigova and D. DeMille, Phys. Rev. A {\bf 82},  063421  (2010).

\bibitem{Ye27062008}
J. Ye, H.~J. Kimble, and H. Katori, Science {\bf 320},  1734  (2008).

\bibitem{PhysRevX.3.031015}
J. Schachenmayer, B.~P. Lanyon, C.~F. Roos, and A.~J. Daley, Phys. Rev. X {\bf
  3},  031015  (2013).

\bibitem{PhysRevLett.111.207202}
P. Hauke and L. Tagliacozzo, Phys. Rev. Lett. {\bf 111},  207202  (2013).

\bibitem{PhysRevLett.111.260401}
J. Eisert, M. van~den Worm, S.~R. Manmana, and M. Kastner, Phys. Rev. Lett.
  {\bf 111},  260401  (2013).

\bibitem{1401.5088}
P. Richerme, Z.-X. Gong, A. Lee, C. Senko, J. Smith, M. Foss-Feig, S.
  Michalakis, A.~V. Gorshkov, and C. Monroe, arXiv:1401.5088  .

\bibitem{1401.5387}
P. Jurcevic, B.~P. Lanyon, P. Hauke, C. Hempel, P. Zoller, R. Blatt, and C.~F.
  Roos, arXiv:1401.5387  .

\bibitem{Guenter}
G. G\"unter, H. Schempp, M. Robert-de Saint-Vincent, V. Gavryusev, S. Helmrich,
  C.~S. Hofmann, S. Whitlock, and M. Weidem\"uller, Science {\bf 342},  954
  (2013).

\bibitem{Chandran_etal}
A. Chandran, V. Khemani, C.~R. Laumann, and S.~L. Sondhi, Phys. Rev. B {\bf
  89},  144201  (2014).

\bibitem{doi:10.1021/cr2003568}
M.~A. Baranov, M. Dalmonte, G. Pupillo, and P. Zoller, Chemical Reviews {\bf
  112},  5012  (2012).

\bibitem{RevModPhys.82.277}
J. Eisert, M. Cramer, and M.~B. Plenio, Rev. Mod. Phys. {\bf 82},  277  (2010).

\bibitem{PRA_72_063407}
X.-L. Deng, D. Porras, and J.~I. Cirac, Phys. Rev. A {\bf 72},  063407  (2005).

\bibitem{PRL_97_150404}
J. Eisert and T.~J. Osborne, Phys. Rev. Lett. {\bf 97},  150404  (2006).

\bibitem{Hauke}
P. Hauke, F.~M. Cucchietti, A. M{\"u}ller-Hermes, M.-C. Ba{\~n}uls, J.~I.
  Cirac, and M. Lewenstein, New J. Phys. {\bf 12},  113037  (2010).

\bibitem{PRL_109_267203}
T. Koffel, M. Lewenstein, and L. Tagliacozzo, Phys. Rev. Lett. {\bf 109},
  267203  (2012).

\bibitem{LRKitaev}
D. Vodola, L. Lepori, E. Ercolessi, A.~V. Gorshkov, and G. Pupillo,
  arXiv:1405.5440  (2014).

\bibitem{PhysRevA.88.023605}
M.~L. Wall, E. Bekaroglu, and L.~D. Carr, Phys. Rev. A {\bf 88},  023605
  (2013).

\bibitem{Brennen2}
A. Micheli, G.~K. Brennen, and P. Zoller, Nature Physics {\bf 2},  341  (2006).

\bibitem{Brennen2007}
G.~K. Brennen, A. Micheli, and P. Zoller, New J. Phys. {\bf 9},  138  (2007).

\bibitem{PhysRevLett.103.155302}
N.~R. Cooper and G.~V. Shlyapnikov, Phys. Rev. Lett. {\bf 103},  155302
  (2009).

\bibitem{PhysRevB.87.081106}
S.~R. Manmana, E.~M. Stoudenmire, K.~R.~A. Hazzard, A.~M. Rey, and A.~V.
  Gorshkov, Phys. Rev. B {\bf 87},  081106  (2013).

\bibitem{doi:10.1080/00268976.2013.800604}
A.~V. Gorshkov, K.~R. Hazzard, and A.~M. Rey, Molecular Physics {\bf 111},
  1908  (2013).

\bibitem{Chiron}
S.~V. Syzranov, M.~L. Wall, V. Gurarie, and A.~M. Rey, arXiv:1406.0570  .

\bibitem{PhysRevLett.109.266804}
N.~Y. Yao, C.~R. Laumann, A.~V. Gorshkov, S.~D. Bennett, E. Demler, P. Zoller,
  and M.~D. Lukin, Phys. Rev. Lett. {\bf 109},  266804  (2012).

\bibitem{PhysRevLett.110.185302}
N.~Y. Yao, A.~V. Gorshkov, C.~R. Laumann, A.~M. L\"auchli, J. Ye, and M.~D.
  Lukin, Phys. Rev. Lett. {\bf 110},  185302  (2013).

\bibitem{wall2013simulating}
M.~L. Wall, K. Maeda, and L.~D. Carr, Annalen der Physik {\bf 525},  845
  (2013).

\bibitem{stuhl2012evaporative}
B.~K. Stuhl, M.~T. Hummon, M. Yeo, G. Qu{\'e}m{\'e}ner, J.~L. Bohn, and J. Ye,
  Nature {\bf 492},  396  (2012).

\end{thebibliography}

\end{document}